\documentclass[conference]{IEEEtran}
\IEEEoverridecommandlockouts
\AtBeginDocument{%
  \providecommand\BibTeX{{%
    \normalfont B\kern-0.5em{\scshape i\kern-0.25em b}\kern-0.8em\TeX}}}

\usepackage{color,soul}
\usepackage{algorithm,amssymb}
\usepackage{algpseudocode}
\usepackage{amsmath,amsthm}

\newtheorem{theorem}{Theorem}
\usepackage{subcaption}
\PassOptionsToPackage{breaklinks = true,unicode = true,urlcolor = blue,colorlinks = true,citecolor = red,linkcolor = red}{hyperref}
\usepackage{hyperref}

\usepackage{bm}

\usepackage[nameinlink,capitalize]{cleveref} 

\usepackage[font=small]{caption}

\newcommand{\paragraphb}[1]{\smallskip\noindent{\bf #1.} }

\usepackage{multirow}
\usepackage{array}
\newcolumntype{P}[1]{>{\centering\arraybackslash}p{#1}} 

\usepackage{enumitem}

\usepackage[nameinlink,capitalize]{cleveref} 

\newcounter{mynote}[section]

\newcommand{\thenote}{\thesection.\arabic{mynote}}

\newcommand{\wbnote}[1]{\ifx\outforreview\undefined\refstepcounter{mynote}{\it\textcolor{blue}{(WB~\thenote: { #1})}}\fi}

\newcommand{\vbnote}[1]{\ifx\outforreview\undefined\refstepcounter{mynote}{\it\textcolor{magenta}{(VB~\thenote: { #1})}}\fi}

\newcommand{\lbnote}[1]{\ifx\outforreview\undefined\refstepcounter{mynote}{\it\textcolor{red}{(LB~\thenote: { #1})}}\fi}

\newcommand{\fixme}[1]{\ifx\outforreview\undefined\textbf{\textcolor{red}{[FIXME: #1]}}\fi}

\newcommand{\pr}[1]{\ensuremath{\mathrm{Pr}(#1)}}
\newcommand{\prd}[2]{\ensuremath{\mathrm{Pr}_{#1}(#2)}}

\newcommand{\mc}[1]{\ensuremath{\mathcal{#1}}}

\newcommand{\transf}{\ensuremath{{\rm Emb}}}

\newcommand{\dist}{\ensuremath{\mc{D}}}
\newcommand{\stdgauss}{\ensuremath{\mc{N}(0,1)}}
\newcommand{\stdgausssigma}{\ensuremath{\mc{N}(0,\sigma^2)}}

\newcommand{\xts}[1]{\ensuremath{{{\bm{x}}_{#1}}}}
\newcommand{\xt}{\xts{T}}
\newcommand{\xzero}{\xts{0}}

\newcommand{\xti}[1]{\ensuremath{{{x}_{T,{#1}}}}}

\usepackage{graphicx}
\usepackage{gensymb}
\usepackage{booktabs}

\begin{document}

\title{Provably Secure Covert Messaging Using Image-based Diffusion Processes}

\author{\IEEEauthorblockN{1\textsuperscript{st} Given Name Surname}
\IEEEauthorblockA{\textit{dept. name of organization (of Aff.)} \\
\textit{name of organization (of Aff.)}\\
City, Country \\
email address or ORCID}
\and
\IEEEauthorblockN{2\textsuperscript{nd} Given Name Surname}
\IEEEauthorblockA{\textit{dept. name of organization (of Aff.)} \\
\textit{name of organization (of Aff.)}\\
City, Country \\
email address or ORCID}}
\author{
    \IEEEauthorblockN{Luke A. Bauer, Wenxuan Bao, and Vincent Bindschaedler}
    \IEEEauthorblockA{University of Florida
    \\\{lukedrebauer, wenxuanbao, vbindschaedler\}@ufl.edu}
}

\maketitle

\begin{abstract}
We consider the problem of securely and robustly embedding covert messages into an image-based diffusion model's output. The sender and receiver want to exchange the maximum amount of information possible per diffusion sampled image while remaining undetected. The adversary wants to detect that such communication is taking place by identifying those diffusion samples that contain covert messages. To maximize robustness to transformations of the diffusion sample, a strategy is for the sender and the receiver to embed the message in the initial latents. We first show that prior work that attempted this is easily broken because their embedding technique alters the latents' distribution. 
We then propose a straightforward method to embed covert messages in the initial latent {\em without} altering the distribution. We prove that our construction achieves indistinguishability to any probabilistic polynomial time adversary. Finally, we discuss and analyze empirically the tradeoffs between embedding capacity, message recovery rates, and robustness. We find that optimizing the inversion method for error correction is crucial for reliability.
\end{abstract}


\begin{figure*}[!t]
     \centering
     \includegraphics[width=.975\linewidth]{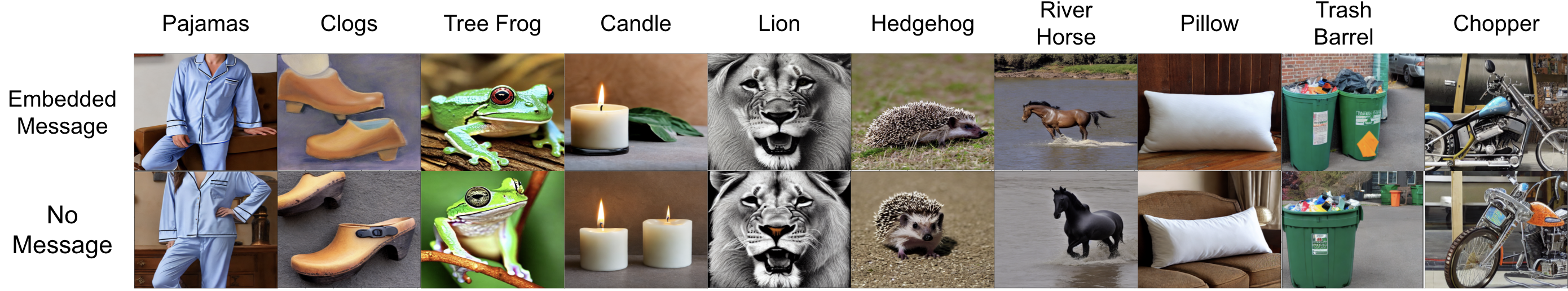}
     
     \caption{Sample images generated using our embedding method and the EDICT scheduler. The top of each column is the prompt. The first row of images contain an embedded message, and the second row contain the same original latent space, without an embedded message.}
     \label{fig:samples}
 \end{figure*}
\section{Introduction}
{\let\thefootnote\relax\footnote{{This work
has been accepted for publication in the IEEE Conference on Secure and
Trustworthy Machine Learning (SaTML). The final version will be
available on IEEE Xplore.}}}

Encrypted communication allow a sender and a receiver to exchange messages securely. These methods, especially those available nowadays, can resist even the most resourceful adversaries' attempts at recovering the confidential messages, but they do {\em not} hide the fact that encrypted communication is taking place. Steganography aims to conceal exchanged messages in innocent looking covers, so that adversaries are unaware that communication is taking place. Traditional steganographic methods achieve this by adding distortion to an existing cover, but such distortion is detectable when the cover is public knowledge and can be compared against. Generative models, especially deep generative models, provide a way to avoid distorting an existing cover, by transforming covert messages into seemingly unique content.

We focus on diffusion models~\cite{sohl2015deep,song2019generative,ho2020denoising,song2020denoising}, which have recently emerged as state-of-the-art generative models for computer vision, due to reportedly outperforming generative adversarial networks (GANs)~\cite{goodfellow2020generative} in output quality~\cite{dhariwal2021diffusion}. Like GANs, we can view diffusion models as mapping a noise distribution into a desired distribution (e.g., images). However, sampling in diffusion models works by iterative denoising. Starting from a pure noise sample, the model transforms it into a realistic output through consecutive denoising steps.

Consider a scenario where Alice wants to send a message to Bob using a diffusion model's outputs as a covert communication channel. 
Concretely, Alice runs the diffusion sampling process based on the bits of her message to Bob so that the output (secretly) embeds those bits. This output image is then sent to Bob indirectly, for example by posting it to some public Internet platform or forum. Bob retrieves the image and then inverts the sampling process using the diffusion model to recover Alice's message. Such a method emphasizes security and concealment over anything else, and as such is ideal for exchanging short, but important messages securely. 

There are a number of evident challenges to overcome to achieve this vision. The first challenge is to efficiently and invertibly encode bits into diffusion outputs. The second challenge is to ensure that the embedding process is robust to transformations of the output images. Otherwise, any modifications (such as those applied by websites when displaying an image) would destroy the covert message.

Furthermore, arguably the most important challenge is ensuring security, i.e., ensuring that an adversary cannot detect whether an output image contains an embedded message. The prevalence of AI/synthetic generated media online means that identifying an output as model-generated is not by itself a tell that it embeds a covert message. However, subtle changes in diffusion process due to the embedding method could be detectable by an adversary, especially since it is reasonable to assume the adversary has access to the same diffusion model as Alice and Bob.

In this paper, we propose a construction that embeds bits in the initial latents to maximize robustness and is compatible with off-the-shelf latent diffusion models such as StableDiffusion. More importantly, our scheme does not alter the latent distribution. We show how this property grants our method indistinguishability at every step of the process, and thus the embedded message in a diffusion sample is undetectable by any polynomial time adversary. 

Our proposed construction has two other desirable features. First, it incorporates a novel error correction technique. This error correction mechanism is built on a specific inversion technique called EDICT~\cite{wallace2023edict}, which performs inversion using dual latent spaces. We show how to leverage the dual latent spaces to obtain improved message recovery rates. Second, we show that embedding covert bits directly in the initial latents noise provides superior robustness to output image transformations (e.g., resize, blur, etc.) than alternatives.

Our paper's contribution is perhaps best conceptualized in relation to the growing literature consisting of numerous prior attempts at doing steganography and (or) covert messaging with generative models such as with text/LLM~\cite{bauer2021covert,kaptchuk2021meteor,cao2022generative,dai2019towards}, speech~\cite{dutta2020overview,li2023coverless,chen2021distribution}, images GANs~\cite{kim2023diffusion,wei2022generative,zhou2022secret,zhu2017unpaired, qin2019coverless,liu2017coverless,zhang2020generative}. However, existing proposals using diffusion models~\cite{peng2023stegaddpm,yu2024cross,wei2023generative,kim2023diffusion} lack robustness, require costly model retraining, and (or) are simply insecure. 

Our main argument in this paper is that previous attempts have failed to properly conceptualize security. Instead of considering the various ways that an adversary could attempt to detect if steganographic communication is taking place, most related work simply evaluate the quality of the model under steganographic embedding, concluding that if output quality is high enough then the system is secure. We show that this is a fallacy. Realistic adversaries are {\em not} limited to examining diffusion outputs. They can use the same inversion process that receivers rely on to recover messages, and use that to identify if steganographic bits are present. 

To demonstrate this, we focus on related work Kim et al.~\cite{kim2023diffusion}, the closest prior attempt to ours in terms of embedding strategy. They propose an embedding strategy they call {\em message projection}. 
They achieve the high output quality they strive for, but distort the latent noise distribution. We construct an attack that breaks their scheme, despite them achieving the standard steganographic ideal of output indistinguishability.

After describing our construction, we empirically evaluate it and analyze its various tradeoffs, with a focus on capacity, message recovery rates, and robustness. We also show through empirical security evaluation (to complement our security proof) that our scheme is undetectable.

\paragraphb{Summary of contributions}
We propose a construction with the following properties. 
\begin{itemize}[noitemsep,nolistsep,leftmargin=1.5em]
    \item {\bf Provable security:} we formalize the notion of {\em latent-space indistinguishability} and prove that our construction achieves it. Unlike related work which only achieves {\em output indistinguishability} (empirically), latent-space indistinguishability is a stronger notion and under it no adversary can identify that covert messaging is taking place.
    
    \item {\bf Distribution-preserving embedding:} we propose a construction that embeds covert messages in the initial latents {\em without} altering the distribution and prove its indistinguishability. 

    \item {\bf Error correction:} we devise an error correction scheme using EDICT~\cite{wallace2023edict}. This improves embedding capacity and increase robustness over the traditional DDIM scheduler.

    \item {\bf Robustness under image transformations:} we analyze robustness of our scheme to various image transformations (e.g., JPEG compression, resizing, etc.) and show that embedding covert messages in the initial latents provides greater robustness than alternatives.
\end{itemize}

\section{Background \& Related works}
\subsection{Diffusion models}

Diffusion models~\cite{sohl2015deep,song2019generative,ho2020denoising,song2020denoising} can be described through their forward and reverse processes. In the forward process, a clean image $\xzero$ is progressively corrupted by adding Gaussian noise through a series of $T$ steps from $\xzero$ to $\xt$. The reverse process aims to reconstruct the original image $\xzero$ by iteratively denoising the noisy image $\xt$. Starting from $\xt$, the model denoises the image step-by-step, reducing the noise and eventually reconstructing the image at $\xzero$. 

\paragraphb{Forward \& Reverse Processes}
Among the various sampling techniques used in diffusion models, such as Denoising Diffusion Probabilistic Models (DDPM)~\cite{ho2020denoising} and Denoising Diffusion Implicit Models (DDIM) \cite{song2020denoising}, deterministic samplers like DDIM are of particular interest. Deterministic samplers are preferred because they are invertible, allowing the reverse process to be precisely defined and controlled. The invertibility of DDIM is crucial for ensuring that the denoising steps can be accurately traced back, facilitating high-fidelity image generation. 
The formula of the forward process in DDIM is:
\[
x_t = \sqrt{\alpha_t} x_0 + \sqrt{1-\alpha_t} \epsilon
\]
where $\alpha_t$ is a predetermined parameter, and $\epsilon$ is a noise vector sampled from the standard normal distribution.
The backward process is given by:
\begin{align*}
x_{t-1} &= \sqrt{\alpha_{t-1}} \left( \frac{x_t - \sqrt{1-\alpha_t} \epsilon_\theta(x_t, t)}{\sqrt{\alpha_t}} \right) \\
&+ \sqrt{1-\alpha_{t-1} - \sigma_t^2} \epsilon_\theta(x_t, t) + \sigma_t z
\end{align*}
where $ \epsilon_\theta(x_t, t)$ is the noise estimated by the model, intended to reverse the diffusion process,
$ \sigma_t $ is an additional noise scale parameter, and
$z$ is random noise sampled from a Gaussian distribution.

In addition to DDIM, we use EDICT~\cite{wallace2023edict}. EDICT is a diffusion scheduler that follows a similar structure to DDIM, but creates two latent spaces. These are initially the same space but diverge during the denoising process as they are continually mixed with each other. This uses a similar estimation process that uses mixing to reduce inversion error, especially for models that use complex prompts.

\paragraphb{Latent Diffusion Models (LDM)}
StableDiffusion\footnote{\url{https://stability.ai/} \& \url{https://huggingface.co/stabilityai/stable-diffusion-2-1}}~\cite{rombach2022high}, an advanced application of diffusion models, operates as a Latent Diffusion Model (LDM). LDMs leverage a latent space representation that the diffusion process operates over. The result is that it enhances the model's ability to generate detailed and high-resolution images. The key components of StableDiffusion include not only the diffusion process over a compressed 64x64 latent space but also an encoder and decoder to convert the latent representations back and forth between the full-size images, typically at resolutions such as 512x512. The encoder compresses the high-resolution images into the latent space, which the diffusion process then modifies. Following this, the decoder reconstructs the modified latent image back into a high-resolution output. The stability and efficiency of LDMs make them a preferred choice for complex image-generation tasks.

\paragraphb{Sampling and Inversion}

In the sampling process, we start with a text prompt $t$ and an initial latent noise sample $\xt$ (isotropic standard Gaussian~\cite{ho2020denoising}), which is represented as a 64x64x4 tensor where the last dimension is the number of channels (RGBA). The StableDiffusion sampling process will then iteratively denoise $\xt$ into $\xzero$ (same dimension), and then a full resolution 512x512 image will be decoded from $\xzero$. Note that this process is completely deterministic once $\xt$ and the prompt are determined. Given a full size resolution 512x512 image (which may or may not have been created through StableDiffusion) the inversion process first uses the encoder to obtain $\tilde{\xzero}$ a 64x64x4 tensor representation of the image. It then iteratively noises this representation back into a latent point $\tilde{\xt}$. As mentioned before since the inversion process is only approximate, the inversion of a sampled initial latent will have lost information. In other words, $\xt \neq \tilde{\xt}$ although we expect $\xt \approx \tilde{\xt}$.

\subsection{Generative Steganography}

Since the advent of deep learning there has been significant interest in doing steganography using deep generative models such as LLMs for text-based steganography~\cite{dai2019towards,kaptchuk2021meteor,bauer2021covert,cao2022generative}, GANs for images~\cite{zhang2019steganogan,hu2018novel,liu2020recent}, and audio/speech~\cite{yang2020approaching,ye2019heard}. In this method, the receiver extracts the message from the generated ``cover media'' through an inversion process or a separate model.

Several researchers have explored image steganography using neural network frameworks such as GAN~\cite{wei2022generative} or Glow~\cite{zhou2022secret}. But were limited by image quality constraints.

Diffusion models have surpassed these other models. Jois et al.~\cite{jois2023pulsar} propose a scheme called ``Pulsar'' for embedding secret data in diffusion model outputs, however, it is designed for pixel space diffusion models, not the more popular latent space models.

There are several schemes using latent space diffusion models, but their approaches vary.  Peng et al.~\cite{peng2023stegaddpm} developed ``StegaDDPM'', embedding secret messages via denoising diffusion models. But this method suffers from minor distortions leading to inversion failures. Yu et al.~\cite{yu2024cross} propose a method called ``CRoSS'', a steganography framework using DDIM to convert cover to secret images, differing from traditional embedding methods. Wei et al.~\cite{wei2023generative} propose Generative Steganography Diffusion, using an invertible model and DCT for data embedding. However, their approach requires significant resources to retrain the model. Peng et al.~\cite{peng2024ldstega} proposed ``LDStega,'' a robust steganography technique for embedding confidential data within images. They demonstrate its resilience to various image distortions, including Gaussian noise and compression. Concurrent work by Mahfuz et al.~\cite{mahfuz2025psyduck} is PSyDUCK. This method embeds a message by diverging the generation process at certain timesteps in the denoising process; it then mixes the diverged samples back together using the covert message. Observing the final image, the receiver can identify the message used to mix the divergent paths. They apply this method to pixel-based and latent space diffusion models for images and videos.

In contrast to existing techniques, our method both uses off-the-shelf latent diffusion models without retraining, and it embeds covert messages in the initial latents (``latent noise'') rather than later on in the diffusion process to obtain greater robustness to modification of the output (e.g., resize, compression, etc.).

Given these desiderata, the work most closely related to ours is Kim et al.~\cite{kim2023diffusion}. They introduce three distinct message projection methods, each designed to embed secret messages into latent noise. The major limitation of their technique is that it alters the latent space distribution. This may not seem to be a significant issue as their outputs appear realistic. But in fact (as we show empirically) it leads to a total break. An adversary can efficiently detect their cover messages with probability almost 1. By contrast, our proposed technique exactly preserves the latent space distribution and we prove it achieves indistinguishability.

\cref{tab:related_work} provides an overview of related work in comparison to our approach. In particular, our paper is the only construction that achieves provable security and proposes an error correction scheme.
\begin{table*}[th]
\centering
\caption{Comparison of our approach with related works.}
\label{tab:related_work}
\resizebox{0.975\textwidth}{!}{%
\begin{tabular}{c|cccccc}
\toprule
Paper  & Retraining Not Required & Robust to Transformations & Embedding in Initial Latents & Preserves Latent Distribution & Error Correction & Provable Security \\ \midrule

Peng et al.\cite{peng2023stegaddpm} & \checkmark  & $\times$  &$\times$ & \checkmark &$\times$ & $\times$ \\
Yu et al.\cite{yu2024cross}          & \checkmark & \checkmark &$\times$  &$\times$  &$\times$ &$\times$  \\
Wei et al.\cite{wei2023generative}  & $\times$ & $\times$ & \checkmark &$\times$  &$\times$ &$\times$  \\
Peng et al. \cite{peng2024ldstega}   &\checkmark  &\checkmark  &$\times$  & $\times$ &$\times$ & $\times$ \\
Kim et al.~\cite{kim2023diffusion}   &\checkmark  & $\times$ &\checkmark
&$\times$ &$\times$  &$\times$  \\
This Paper &\checkmark  &\checkmark   &\checkmark  &\checkmark  &\checkmark &\checkmark \\ \bottomrule
\end{tabular}%
}
\end{table*}

\subsection{Watermarking}
A different problem that relies on similar techniques to ours is watermarking; in which a model owner embeds a small watermark within the generated diffusion output. This problem has received substantial attention recently~\cite{wen2023tree,yang2024gaussian,liu2023watermarking,zhao2023recipe}. In particular, two methods also attempt to invisibly embed information in the latent space. Tree Ring Watermarking~\cite{wen2023tree} embeds in patterns structured in Fourier space that are highly robust, but influence model output. Gaussian Shading~\cite{yang2024gaussian} seeks to match the natural Gaussian distribution, and achieves high robustness, but has limited capacity. We discuss these papers because watermarking has similarities to generative steganography in that both attempt to embed information into generated images without decreasing the quality of the output. However, they differ in their goals and threat models. Watermarking strives for easy retrieval and robustness against alterations. By contrast, the goal of steganography is to create a reliable channel between the sender and receiver that is completely undetectable and maximizes its capacity.

\section{Problem Statement \& Threat Model}
We consider a sender and a receiver communicating covertly by using diffusion-generated images. The sender, Alice, embeds a plaintext message into the generation process of a (public) diffusion model. This process results in images that are then sent to Bob, the intended receiver. Bob then proceeds to invert the diffusion process to recover the message. 

In this paper, we are not concerned with how the diffusion output images are sent to Bob. For example, they may be shared over a public platform such as Reddit or X. Or in some cases Alice and Bob may in fact be the same individual and the covert messaging channel created is thus a kind of plausibly deniable storage scheme that Alice can use (e.g., Alice has to surrender her personal computing devices at a border crossing to/from an authoritarian regime). 

Regardless of specifics, the scenarios we consider prioritize security and reliability over transmission capacity. Therefore, the construction we propose is not designed for frequent low-latency large capacity communication, such as private messaging, but is intended for covert communication (of small payloads) that must not be discovered. 

To ensure confidentiality, integrity, and authentication, the message embedded by Alice will first be encrypted and authenticated using a shared key with Bob. We assume Alice and Bob have already shared a cryptographic key pair before initiating any communication.

\begin{figure*}[!t]
     \centering
     \includegraphics[width=\linewidth]{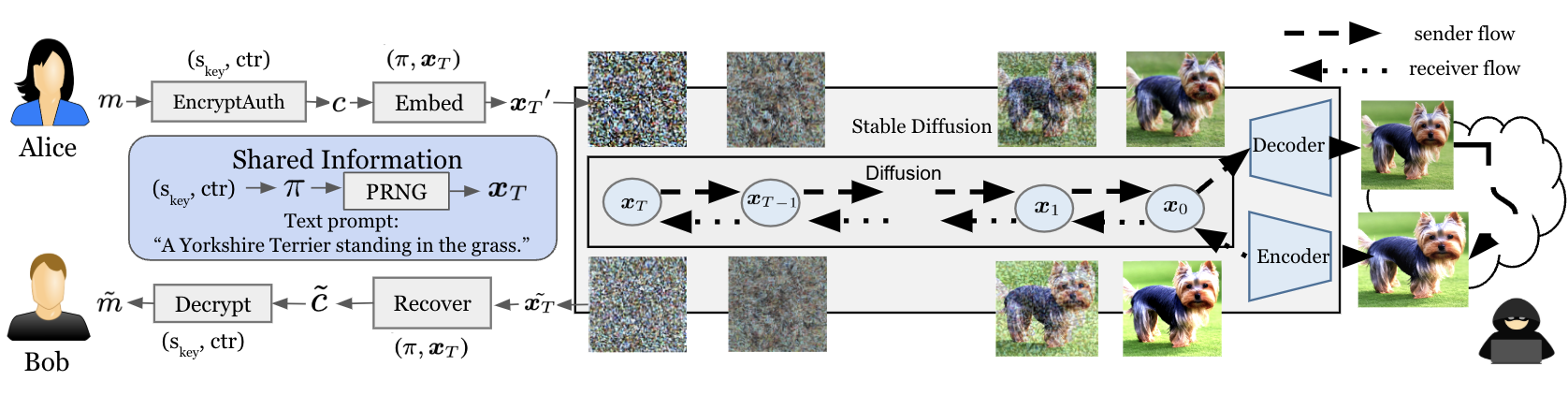}
     
     \caption{Proposed construction sender-side and receiver-side operations. Shared information must be agreed upon by both the sender and receiver, but much of it can be stored in the parameter table. Alice encrypts then embeds the message into the latent space which is used to generate the cover image. The cover is then received by Bob, the ciphertext is recovered and then decrypted giving Bob the original message. }
     \label{fig:overview}

 \end{figure*}

\subsection{Threat Model}\label{sec:threatmodel}
We assume an adversary whose goal is to identify that covert communication is taking place and (if so) who are the communicating senders and receivers. The adversary's goal is therefore to distinguish between diffusion-generated images that contain covert messages and those that do not.

The adversary has access to every aspect of our proposed system, except for the shared secret key between the sender and the receiver. This means the adversary has full knowledge of/access to the diffusion model, hyperparameters used, details of the cryptographic algorithms used, prompts, etc. In particular the adversary can generate their own outputs from the diffusion model, as well as attempt to invert any image back into (reconstructed) latents. However, the adversary does not observe the initial randomness chosen in the sender's latents. We also assume that the adversary runs in probabilistic polynomial time (PPT) in the security parameter (in our case the size in bits of the shared key materials). This is because otherwise the adversary could break the cryptography to recover the plaintext directly (ignoring how the diffusion-based generation works).  

We emphasize that the adversary can invert an image using the diffusion model, retrieving the same latent space as the receiver. In other words, the adversary can act as a sender/receiver attempting to decode messages as well as sending fake ones. 

Moreover, the adversary can train a (PPT) distinguisher and query the model to identify its distribution, as well as the distribution of the noise used in the image generation process. 

We consider the possibility that diffusion-generated images will be altered by image transformations such as resizing or format transformations (e.g., by an adversary or by the platform). This is mostly to reflect the fact many platforms perform such transformations when attempting to display images in certain sizes or shapes, such as for a user portrait or when compressing the image to reduce load on the site.
This is not meant to capture an adversary that seeks to completely prevent covert messages, as such an adversary could use image transformation that would completely destroy any communication (e.g., resizing the image to 1x1 pixel, replacing it with a black square, etc.). We also do not consider an adversary that seeks to identify steganographic posts based on the subjects of the post, or the behavior of the user.

We do not consider the cases where the adversary controls the process by which the sender and receiver exchange images (e.g, full control of the platform images are posted on). In which case the adversary could perform other attacks, such as deleting images, replacing images with adversarial examples, analyzing posting habits, and observing scraping activity. We leave such analysis to future work.

\subsection{Security Desiderata \& Other Goals}

Within the scenario described above there are several goals we wish to achieve. 
\begin{itemize}[leftmargin=1.5em,noitemsep,nolistsep]
    \item \textbf{Covertness:} Exchange messages without the adversary being aware an exchange is taking place. 
    \item \textbf{Output Indistinguishability:} It should not be possible to distinguish a steganographic generated image from a non-steganographic generated image, without the secret key.
    \item \textbf{Latent Space Indistinguishability:} If the adversary were to attempt to invert a non-steganographic image and a steganographic one, the returned latent spaces should follow the same distribution and be indistinguishable. 
    \item \textbf{Confidentiality and Integrity:} The only person who should be able to identify that a covert message is embedded within the noise should be Bob, who uses his secret key to retrieve the message. Furthermore, Bob should be able to determine if the message was altered.
\end{itemize}

In addition, the steganographic method should be useful for covert messaging. This requires adding few goals relating to performance. 
\begin{itemize}[leftmargin=1.5em,noitemsep,nolistsep]
    \item \textbf{Utility:} Alice must be able to embed enough bits per diffusion output for transmitting useful messages, and Bob must be able to reliably extract the messages (with reasonably high probability of correctness).
    \item \textbf{Robustness:} The embedding method should be robust enough to retrieve the message after enduring transformations common on public sites, such as quality reduction, compression, and format changes.
\end{itemize}

If the adversary alters the images Bob may not be able to retrieve them. However, this should not result in the adversary being able to identify a steganographic image, nor should it result in an image Bob could mistake for the original.

\section{Proposed Construction}\label{sec:construction}\label{sec:scheme}\label{sec:method}

Our method can be described in four parts: (1) setup; (2) cryptographic record; (3) embedding; and (4) retrieval. \cref{fig:overview} provides an overview of our construction.

\subsection{Initial Setup}

In generative steganography, the sender and receiver must use the same model with the same prompt and parameters, as well as the same steganographic settings. Otherwise they will observe different distributions and be unable to exchange information. 
We assume the sender and receiver have exchanged a secret key $s_{\rm key}$, which the adversary does not observe, as well as a message counter ${\rm ctr}$ for the number of messages exchanged using this key. From this they can then derive unique key materials for each message from this key and message counter. We use this key material to seed PRNGs that are used to generate the initial latent $\xt$, permutation $\pi$, and select a row in the parameters table.

The parameter table associates sets of parameters together so that the sender and receiver can stay synchronized in their use of parameters. Each row in the table includes information such as the model name/version, the prompt, the threshold and redundancy levels, and any other information needed to synchronize states. This allows the sender and receiver to exchange multiple messages with varying parameters with only a one time exchange of the secret key $s_{\rm key}$. The parameter table is ideally private, but that may not be realistic, for logistical and adversarial reasons, so we can assume the table is public. As such the sender and receiver could download it from a public location, ensuring they have the same model and parameter settings. Consistent with our threat model, this means the adversary also has access to this table, and has knowledge of all models and parameters. We assume the models used by the sender and receiver are versions publicly available on sites that host models/finetunings of image generation models. This is not a problem, because security relies only on the secrecy of the key. 

It is worth emphasizing that we need not assume anything about the model or its training set. Moreover, if the adversary were somehow able to corrupt the model so that the sender and receiver downloaded different versions, that could prevent correct exchange of messages, but would not undermine the security of the scheme. If this is a concern the sender and receiver can use hash/checksums to ensure integrity of their local copy of the model.

For simplicity of exposition, we sometimes omit embedding parameters or hyperparameters as input to the algorithms. But we assume both sender and receiver have agreed upon a set of values to use.

\begin{algorithm}[!t]
  \caption{Embed Ciphertext into Latent Space} 
  \label{alg:embed}
  \small
  \begin{algorithmic}[1]
    \Require

      $c$: Ciphertext; 

      $\pi$: Permutation;

      $\xt$: Initial latents;

       $\rho$: Redundancy parameter;

       $\tau$: Threshold parameter;

    \Ensure
    $\xt'$: Latents with embedded ciphertext $c$
  
   \State $c' \gets {\rm AddRedundancy}(c,\rho)$  \Comment{\begin{scriptsize} Within $c$ convert all 0 bits to -1, then apply redundancy by repeating each bit $\rho$ times.\end{scriptsize}}

    \State ${\rm c_m'} \gets {\rm Permute}(c', \pi)$ \Comment{\begin{scriptsize} Permute $c'$,  according to $\pi$ \end{scriptsize}}
    \State $ \xt' \gets \xt$ 
    \For {$\xti{i}$ in $\xt'$} \Comment{\begin{scriptsize}  Loop over each latent component in $\xt'$\end{scriptsize}}
        \If{$|\xti{i}| \geq \tau$} \Comment{\begin{scriptsize}  Only embed in values larger magnitude than $\tau$\end{scriptsize}} 
        \State $s \gets c'_m.{\rm pop}()$
            \If{$s = -1$}
                \State $\xti{i} \gets - \xti{i}$ \Comment{\begin{scriptsize} Flip the sign of $i^{\rm th}$ element of $\xt'$ \end{scriptsize}}
            \EndIf      
        
        \EndIf
    \EndFor

    \State\Return{$\xt'$}

  \end{algorithmic}
\end{algorithm}
\subsection{Cryptographic Record}
Before Alice can send the plaintext message $m$ to Bob, she first must encrypt it and ensure integrity and authenticity. 

For this, we use an Authenticated Encryption with Associated Data (AEAD) construction realized in practice using an AES block-cipher in CTR-mode to encrypt our plaintext giving us $c_{\rm body}$. The counter allows them to exchange multiple messages using the same key without compromising message security and integrity. We also generate an authentication tag, $c_{\rm tag}$, by computing an SHA512-HMAC of $c_{\rm body}$ using the $s_{\rm key}$ and truncating the result (e.g., to five bytes in experiments). The tag $c_{\rm tag}$ is appended to the end of $c_{\rm body}$ giving us our ciphertext $c = c_{\rm body} || c_{\rm tag}$. The tag can be checked by the receiver on the retrieved message to ensure no bits were lost from inversion or adversarial interference. Both the AES block cipher and the HMAC tag use keys derived from the secret key $s_{\rm key}$ initially exchanged between the sender and receiver.

\subsection{Ciphertext Embedding}

Recall that due to their shared key material, Alice and Bob have agreed on parameters. For StableDiffusion, the initial latents $\xt$ are represented as 64x64x4 tensors sampled as isotropic Gaussian~\cite{ho2020denoising}. Said differently, each latent component $\xti{i}$ is sampled i.i.d. from $\stdgauss$. So the only difference in this case is that the PRNG is seeded based on key material derived from the shared secret key and the message counter. This way Alice and Bob will both obtain the same initial latents $\xt$ for each message and can use a similar process to generate the same permutation $\pi$.

Given this, Alice has to perform two steps for embedding a ciphertext $c = c_{\rm body} || c_{\rm tag}$. 
First, redundancy is added to the ciphertext as follows. Each bit in $c$ is repeated $\rho \geq 1$ times (redundancy parameter) before permuting the (redundant) ciphertext according to permutation $\pi$, creating $c'_m$.  Second, we only embed bits into component values (of $\xt$) with absolute values above a certain threshold $\tau \geq 0$. So for each component value within the 64x64x4 latent space, if its absolute value exceeds $\tau$ we embed a bit from $c'_m$. To embed a $1$ bit, we do not change the component value, and to embed a $0$ bit, we flip the current sign of the component value. As we show in~\cref{sec:security:indistinguishability}, this flipping of bits does not alter the distribution. The method is illustrated in~\cref{alg:embed}. Lines 7-8 perform the actual logic embedding bits. The process can be thought of as applying a mask of permuted values over valid values.
If we have redundancy $\rho$ then we can only embed $16384/\rho$ bits of the ciphertext at the most per output (assuming 64x64x4 latent space), and any component value that does not exceed $\tau$ in magnitude is not embedded with a bit.  We discuss this further in~\cref{sec:perf}.

\begin{algorithm}[!t]
  \caption{Recover Ciphertext} 
  \label{alg:Recover}
  \small
  \begin{algorithmic}[1]
    \Require
      
      $\xt$: Initial latents;

      $\pi$: Permutation;

      $\tilde{\xt}$: Reconstructed latents;
      
      $\rho$: Redundancy parameter;

      $\tau$: Threshold parameter;

      $p(\cdot)$: pdf of $\stdgauss$;  
     \Comment{\begin{scriptsize}  Assume errors are standard normal\end{scriptsize}}

    \Ensure
    $\tilde{c
    }$: reconstructed ciphertext
     \Procedure{Flatten}{$\xt, \tau$}
        \State We wish to convert a 64x64x4 latent space into a single flat list of valid components.
        \State $v \gets []$
        \State Going layer by layer, append to $v$ each latent space component from $\xt$ if their absolute value is greater than $\tau$. 
        \State \Return $v$

    \EndProcedure

    \Procedure{Remove\_Mask}{$\rm v, \pi,\rho, \tau$}
        \State $\rm v' \gets {\rm Inverse\_permute}(\rm v, \pi, \tau)$ \Comment{\begin{scriptsize} Given a list of latent space values remove the applied Mask by identifying all values with a magnitude greater than $\tau$, and then inverting the permutation. Returning latent space values respective to $c'$ ($c$ with redundancy applied) \end{scriptsize}}
        \State $\rm v' \gets {\rm Partition}(v, \rho)$ \Comment{\begin{scriptsize}Partition list into a list of $\rho$ length lists
        such that sub-list $i$ contains all the latent space values embedded with the $i$th bit of $c$\end{scriptsize}}
        \State return $v'$
    \EndProcedure

    \State ${ \rm v_{\rm orig}} \gets {\rm Flatten}(\xt)$

    \State ${ \rm v_{\rm recov}}\gets {\rm Flatten}(\tilde\xt)$

    \State ${\rm v_{orig}'} \gets {\rm Remove\_Mask}({\rm v_{orig}}, \pi, \rho, \tau)$

    \State ${\rm v_{recov}'} \gets {\rm Remove\_Mask}({\rm v_{recov}}, \pi, \rho, \tau)$

    \State $\tilde{c} \gets ``"$

    \For{$i \gets 0$ to len(c)} 
    \State Use~\cref{eq:lr}
    to recover the $i$th bit
    \State Add the bit to $\tilde{c}$

    \EndFor

    \State\Return{$\tilde{c}$}

  \end{algorithmic}
\end{algorithm}

\subsection{Latent Space Inversion}

Our construction uses the DDIM scheduler inversion process proposed by Ho et al.~\cite{ho2020denoising} and as such can be used with any diffusion model using that scheduler. In this paper, in addition to using the DDIM scheduler, we also propose to use EDICT, a variant of DDIM that uses dual latent spaces to improve inversion accuracy \cite{wallace2023edict}. More specifically, when denoising at each step, the scheduler will update one latent space with information from the other one, in a deterministic and invertible way. This helps stabilize the inversion process, allowing us to perform error correction. The use of EDICT increases message accuracy (the percentage of messages sent that can be perfectly recovered) as we detail in~\cref{sec:perf}. When embedding, the only difference between EDICT and DDIM is that we duplicate $\xt'$, to form $(\xt',\xt')$ which functions as the dual latent spaces that initiate image generation. On the receiving end we perform the same retrieval process across both latent spaces, and then use them to perform error correction as detailed in~\cref{alg:correct}.

\subsection{Ciphertext Recovery}

\begin{algorithm}[!t]
  \caption{Error Correction} 
  \label{alg:correct}
  \small
  \begin{algorithmic}[1]
    \Require
      $s_{\rm key}$: secret key;
      
      $\tilde{c_1}$: reconstructed ciphertext of first latents;

      $\tilde{c_2}$: reconstructed ciphertext of second latents;

      ${\rm max\_errs}$: maximum potential errors to correct \Comment{\begin{scriptsize}  default: 10\end{scriptsize}}

    \Ensure
    $m$: message, or Fail
    
    \Procedure{Check}{$c, s_{\rm key}$}
        \State Check tag of ciphertext, 
 if valid return True, else return False
    \EndProcedure
    
    \State ${\rm err} \gets {\rm Compare}(\tilde{c_1}, \tilde{c_2}) $ \Comment{\begin{scriptsize} Identify indexes where the ciphertexts differ. \end{scriptsize}}
   
    \If{${\rm len}({\rm err}) > {\rm max\_errs}$}
        \State \Return Fail            
    \EndIf
    
    \State Use recursion to generate $2^{\rm max\_errs}$ possible ciphertexts.
    \State Each ciphertext being a possible combination of error bit replacements.
    \For{$c$ in $2^{\rm max\_errs}$ possible ciphertexts}
    
        \If{Check($c, s_{\rm key}$)}
             \State \Return Decryption of valid $c$
        \EndIf
        
    \EndFor

   \State \Return Fail
   
  \end{algorithmic}
\end{algorithm}

Once an image has gone through inversion, the receiver has an inverted latents $\tilde{\xt}$ (or $(\tilde{\xt},\tilde{\xt'})$ in the case of EDICT). Recall that the receiver also knows $\xt$, the initial latents used by the sender for this message. Using $\xt$ to identify which component values had magnitudes above the threshold $\tau$, the receiver can compute the ${\rm mask}$ used by the sender and extract the component values associated with it, and then invert the permutation to recover latent components that encode the ciphertext, with redundancy. 

Once the mask is removed, the receiver has a list of component values matching the ciphertext with added redundancy. The receiver then splits the list into ${\rm len}(c)$ lists of length $\rho$. Each list in ${\rm v_{recov}'}$ contains $\rho$ component values, which were embedded with the redundant bits for bit $i$ in the ciphertext. Thus each list ${\rm v_{recov}'}[i]$ can be associated with the $i$-th ciphertext bit. We repeat this process on the initial latents $\xt$, to receive ${\rm v_{orig}'}$, with the same list format but containing component values from $\xt$.

For each bit $i$, the receiver then compares ${\rm v_{orig}'}[i]$ and ${\rm v_{recov}'}[i]$. If all components of each of the two lists have the same sign, then recovering the value of embedding bit $i$ is straightforward. It is $1$ if the signs in ${\rm v_{recov}'}[i]$ match the signs in ${\rm v_{orig}'}[i]$ and 0 otherwise. (Recall that during embedding Alice flips the sign of each component if the corresponding ciphertext bit is 0.)

The challenge is that because the inversion process is not perfect, within the list ${\rm v_{recov}'}[i]$ there may be components of different signs. In that case, we propose to recover the ciphertext bit using a likelihood ratio test based on the assumption that inversion errors are i.i.d. standard normal distributed. Said differently, we assume that if the value of a component (of $\xt$) is initially $x$ then after embedding and inversion it is $x + z$, where $z \sim \stdgausssigma$ for some unknown but fixed $\sigma > 0$.

This strategy allows us to take into account the magnitude of each component value in the recovery (as opposed to only its sign). For example, it is more likely that inversion error caused $0.02$ to flip to $-0.01$ than it is that $3.0$ flipped to $-3.2$. 

Concretely, the likelihood ratio we use to retrieve the $i$th bit is:
\begin{equation}\label{eq:lr}
        \frac{\prod_{j = 0}^{\rho-1} p({\rm v_{orig}'}[i][j]-{\rm v_{recov}'}[i][j])}
        {\prod_{j = 0}^{\rho-1} p(-{\rm v_{orig}'}[i][j] - {\rm v_{recov}'}[i][j])} 
\end{equation}
where $p(\cdot)$ denotes the standard normal probability density function.\footnote{We can use the standard normal PDF for this and do not need to know the common variance $\sigma^2$ since it does not affect whether the ratio is greater than 1.}

If the ratio in~\cref{eq:lr} is greater than 1 then it is more likely (given the standard normal error assumption) that ciphertext bit $i$ was not flipped (so $c_i = 1$). Otherwise, we conclude ciphertext bit $i$ was most likely 0. We use this equation and show the entire process in~\cref{alg:Recover}.

With this, we can recover the ciphertext. We can then verify that we have correctly retrieved the ciphertext by comparing the ${\rm tag}$ at the end of the ciphertext against an HMAC of the body of the ciphertext. If they match we have correctly retrieved the message. If they do not then there has been an error. Note that in addition to ensuring integrity, this also proves the message came from Alice since without the key an adversary cannot forge a valid HMAC tag.

When using the DDIM scheduler, this is the end of the recovery process. If the HMAC tag does not match, then the message is lost. 

\paragraphb{Error Correction with EDICT}
By contrast, when using EDICT we can perform additional error correction. Our novel proposal works as follows. Due to EDICT we have dual latents $(\tilde{\xt},\tilde{\xt'})$ and therefore we can use~\cref{alg:Recover} and~\cref{eq:lr} to create two reconstructed ciphertexts $\tilde{c_1}, \tilde{c_2}$, one for each inverted latents. Both ciphertext can be checked to see if the HMAC tag is valid. But if neither is recovered without error, we can use their differences to identify and recover correct bits. If the number of differing bits between $\tilde{c_1}$ and $\tilde{c_2}$ is small, then we can create a new ciphertext for every combination of differing bits and check if the HMAC tag is valid.

In practice, assuming that reasonable embedding parameters $\rho$ and $\tau$ were used, there is often only a small number of problem bits (e.g., less than $10$) in practice. Therefore we can check every possibility quickly. If we observe $n$ problem bits, then there are $2^n$ possible ciphertext to check. So, we set a maximum number of problem bits ${\rm max\_errs}$ to attempt recovery (e.g., ${\rm max\_errs}=10$) to bound the computational effort. This method allows recovering any message for which there are only a few errors due to inversion, and for which both retrieved ciphertexts do not incorrectly identify the same bit. If an incorrect bit is not among the problem bits, then the recovered ciphertext is incorrect and the message is lost.

\section{Formalizing and Analyzing Security}\label{sec:security}

In this section, we analyze the security of our proposed approach following our threat model, defined in~\cref{sec:threatmodel}.

\paragraphb{Defining Security \& Indistinguishabilities}
Recall that the adversary's proximate goal is to distinguish diffusion images that contain embedded messages from those that do not. Therefore, a straightforward way to define security is as {\em output indistinguishability}, i.e., the scheme is secure if the output images produced by the diffusion-based steganographic scheme are indistinguishable from images produced by the diffusion model itself.  This is an appealing way to define security, in part because it is easy to evaluate empirically. It suffices to generate a dataset of images produced by the steganographic scheme and a dataset of images produced by the diffusion model alone and compare the two, for example, using an ML-based distinguisher. This is what most related works do~\cite{peng2023stegaddpm,yu2024cross,wei2023generative,kim2023diffusion}.

The problem with this reasoning is that the adversary is not limited to only analyzing the diffusion images themselves. The adversary can reason about the process by which images are obtained and even simulate the sender or receiver process. Of course, the adversary can never recover the plaintext message embedded in the diffusion image (without the secret key or breaking the cryptography). But by an approximate inversion of the image they can recover approximate latents $\tilde{\xt}'$, which they may use to infer the presence of an embedded message. Therefore, output indistinguishability is insufficient. A secure scheme needs to ensure that the distribution of any recovered latents $\tilde{\xt'}$ on diffusion output embedding some covert messages matches exactly the distribution of recovered latents on (non-steganographic) outputs. A principled way to do this in our setting is to achieve {\em latent-space indistinguishability}, i.e., ensure that the initial latent-space distributions with and without embedding covert messages are identical. 

Latent-space indistinguishability is a stronger notion. It implies output indistinguishability because once initial latents are sampled, the diffusion process is completely deterministic. If the initial latents with and without covert messages are identically distributed, then the output with and without covert messages will also be identically distributed.\footnote{Once we achieve latent space indistinguishability, we do not need to worry about the output quality of images embedding covert messages; it will be identical to those images that do not embed anything. Nevertheless, we evaluate this empirically in~\cref{sec:experiments:coversecurity}.}

In the rest of this section, we prove the security of our construction. First, we formalize initial latents schemes (\cref{sec:security:lsform}). We then propose an indistinguishability game for them (\cref{sec:security:indgame}). In the process, we explain why our proposed approach does not suffer from the same issue as Kim et al.~\cite{kim2023diffusion} (\cref{sec:security:kimetal}). Finally, we prove our construction's indistinguishability (\cref{sec:security:indistinguishability}).

\subsection{Initial Latents Steganography}\label{sec:security:lsform}
We are given a security parameter $l_{\rm sec}$ and a latent space dimension $k$. A {\em initial latents steganography} scheme is a tuple (Init, Embed, Recover) defined as follows.
\begin{itemize}[leftmargin=1.5em,noitemsep,nolistsep]
    \item Init$(l_{\rm sec})$ produces a secret key $s_{\rm key}$ of $l_{\rm sec}$ bits.
    \item ${\rm Embed}(s_{\rm key}, m, \xt)$: encrypts and embeds a message $m$ of $0 < l_m \leq k$ bits under key $s_{\rm key}$ starting from initial latents $\xt \in \mathbb{R}^k$. The output is some $\xt' \in \mathbb{R}^k$.
    \item ${\rm Recover}(s_{\rm key}, \tilde{\xt}, \xt)$: recovers plaintext from inverted latents $\tilde{\xt} \in \mathbb{R}^k$ and initial latents $\xt$ under key $s_{\rm key}$. The output is a plaintext $\tilde{m}$ of length $l_m$ or $\bot$ (failure).
\end{itemize}

The correctness condition is that if for any tuple $(s_{\rm key}, \xt, m)$, if ${\rm Embed}(s_{\rm key}, m, \xt)$ outputs $\xt'$, then ${\rm Recover}(s_{\rm key}, \xt', \xt)$ outputs $m'$ such that $m=m'$. In other words, assuming zero inversion loss, the message is recovered with probability $1$ under any key and initial latents. In practice, since there will be inversion loss, we are never guaranteed to recover any message.

Observe that the construction does not involve the diffusion model at all. This is by design and a consequence of the fact that we embed bits in the initial latents.

\subsection{Indistinguishability Game}\label{sec:security:indgame}
We propose an indistinguishability game to formalize latent space indistinguishability for diffusion. This is the main security notion we use in this paper.

We assume that the challenger and adversary have common diffusion model $\mc{G}$ with a prompt $t$ and sampling hyperparameters ${\rm params}$. We also have $l_{\rm sec}$ the security parameter in bits and a message length $l_m > 0$.

The game proceeds as follows.
\begin{enumerate}[leftmargin=1.5em,noitemsep,nolistsep]
    \item The challenger generates key material $s_{\rm key}$ of $l_{\rm sec}$ bits and also a uniformly random bit $b \in \{0,1\}$.
    
    \item The adversary generates a plaintext message $m$ of at most $l_m$ bits and sends it to the challenger.
    
    \item The challenger pads $m$ to $l_m$ bits. The challenger then generates $\xt$ from the diffusion model $\mc{G}$ initial latent distribution $\dist$. If $b = 1$, the challenger invokes ${\rm Embed}(s_{\rm key}, m, \xt)$ to obtain $\xt'$ and then sets $\xt = \xt'$. Otherwise (if $b=0$) then the challenger does nothing.
    
    \item The challenger invokes the diffusion model $\mc{G}$ with a prompt $t$ and sampling hyperparameters ${\rm params}$ on $\xt$. It obtains the corresponding image $\xzero$ and sends it to the adversary.
    
    \item The adversary then runs PPT (in $l_{\rm sec}$) algorithm $A(\mc{G}, t, {\rm params}, \xzero)$ to obtain a bit $b'$. The adversary outputs $b'$.
\end{enumerate}

If $b'=b$ then the adversary wins the game. Moreover, we define the adversary success rate as $\pr{b'=b}$ and consequently the adversary's advantage as ${\rm Adv} = 2\pr{b'=b} - 1$. 

We say a scheme is secure if no PPT adversary can win the game with advantage greater than a negligible function with respect to the security parameter $l_{\rm sec}$. In other words, if for every PPT adversary we have: ${\rm Adv} \leq {\rm negl}(l_{\rm sec})$.

\subsection{The Problem with Distorting Distributions}\label{sec:security:kimetal}
Kim et al.~\cite{kim2023diffusion}, our closest prior work, alters the initial latents distribution as part of their embedding process. Indeed, instead of generating a random Gaussian latent space, they use message projection to fill the latent space with values according to the bits they need to embed. So instead of flipping a value's sign (as we propose), they replace it with 1 for one bits, and -1 for zero bits. Alternatively they pull from a small subset of values, i.e. $\sqrt{2}$ or $-\sqrt{2}$ for 1 bits and $0$ for zero bits. 

The result is that after embedding their latents do not follow an isotropic Gaussian distribution. Their method can match the mean and standard deviation of the standard Gaussian distribution, but in reality there only three possible values (0, $\sqrt{2}$, or $-\sqrt{2}$). This discrepancy in distribution can be exploited by an adversary to identify covert messages.

Under realistic threat models (such as ours) the adversary does not observe the latents directly. Instead they observe diffusion outputs from which they can attempt to reconstruct the latents by inversion. As we show in experiments (\cref{sec:exp:kimetaldist}), despite significant loss in the inversion process, the proposal of Kim et al.~\cite{kim2023diffusion} is not secure. An adversary that approximately reconstructs the latent and trains a simple ML-based distinguisher (e.g., a Random Forest classifier) readily achieves 100\% accuracy on distinguishing between their latent spaces and non-steganographic latent spaces. 

Interestingly, despite altering the latents distribution, the construction of Kim et al.~\cite{kim2023diffusion} produces high quality diffusion outputs (as we confirmed experimentally --- \cref{sec:experiments:coversecurity}). The pitfall of their approach is the (implicit) assumption that high output quality implies security. 

\subsection{Achieving Latents Indistinguishability}\label{sec:security:indistinguishability}
Our embedding procedure (\cref{alg:embed}) does not alter the distribution of the initial latents. Intuitively, the reason is that the initial latents are sampled i.i.d. from a symmetric distribution and \cref{alg:embed} simply flips some of the signs based on the redundancy-expanded ciphertext bits. Interestingly, this argument would not hold true if for example the algorithm was to set the sign based on the redundancy-expanded ciphertext bits (e.g., set $\xti{i} \leftarrow +|\xti{i}|$ if bit $i$ is 1 and $\xti{i} \leftarrow -|\xti{i}|$ otherwise). For instance, in such a case, the number of positive entries in the transformed latents would be a multiple of the redundancy. This could then be used by an adversary to distinguish latents that embed bits from those that do not.

The following theorem makes this intuition precise. It states that~\cref{alg:embed} does not alter the natural sampling distribution of the underlying (latent) diffusion model. 
\begin{theorem}\label{thm:embeddist}
    Let $\xt$ be a vector of $k$ random variables each distributed according to a symmetric distribution $\dist$, i.e., $\xti{i} \sim \dist$ for $i = 1,2,\ldots, k$ i.i.d. Also, let $\transf$ denote the transformation of $\xt$ into $\xt'$ in~\cref{alg:embed}. Then the distribution of $\xt$ is not altered by $\transf$, i.e., we have: $\transf(\xti{i}) \sim \dist$ for $i = 1,2,\ldots, k$.
\end{theorem}

\cref{thm:embeddist} states that any adversary (even one not bounded in probabilistic polynomial time) cannot distinguish between initial latents $\xt$ and initial latents $\xt'=\transf(\xt)$ that embed ciphertext bits. Since given the latents, the diffusion process is entirely deterministic (and public) the corollary is that diffusion outputs that embed ciphertext bits are also (statistically) indistinguishable from those that do not. It immediately follows that no PPT adversary can win with higher than negligible advantage the indistinguishability game of~\cref{sec:security:indgame}. That said, this does not imply the entire construction is statistically secure (it is only PPT secure). The reason is that a non PPT adversary could break the cryptography itself, even if they cannot directly distinguish the latents that embed ciphertext bits from those that do not.

\begin{proof}[Proof of \cref{thm:embeddist}]
    By construction,~\cref{alg:embed} applies to each component of $\xt$ independently (based on each bit in ciphertext $c$). So it suffices to consider what happens to $\xti{i}$ for an arbitrary $i$. Let $\prd{\dist}{\xti{i}}$ denote the probability density function of $\xti{i}$ under $\dist$. 
    If the embedded bit is 0 then there is no change ($\xti{i} = \transf(\xti{i})$). If the embedded bit is 1 then $\transf(\xti{i}) = -\xti{i}$. But since $\dist$ is symmetric, i.e., $\prd{\dist}{x} = \prd{\dist}{-x}$ for any $x \in \mathbb{R}$, there is no change in the distribution and we conclude that $\transf(\xti{i}) \sim \dist$.
\end{proof}

Remark that~\cref{thm:embeddist} only requires that the distribution be symmetric and each component be sampled i.i.d. (e.g., it need not be Gaussian). However, in our case with StableDiffusion the distribution of each latent space component is standard normal distributed (i.e., $\dist = \stdgauss$).

\section{Experimental Setup}
As our LDM, we use CompVis/stable-diffusion-v1-4 \cite{Rombach_2022_CVPR} with one of two schedulers, DDIM or EDICT. For prompts we use one of ImageNet's 1000 classes \cite{deng2009imagenet}. For the diffusion process, we use 50 inference steps, and we generate 512x512 images. Unless otherwise stated we use a threshold of 0.3 and a redundancy of 6, and the EDICT scheduler. All experiments are done on servers with 12 GB of RAM for the CPU, and an NVIDIA A100 NVSWITCH GPU with 80 GB of memory. 

\begin{figure}[!t]
     \centering
     \includegraphics[width=0.975\linewidth]{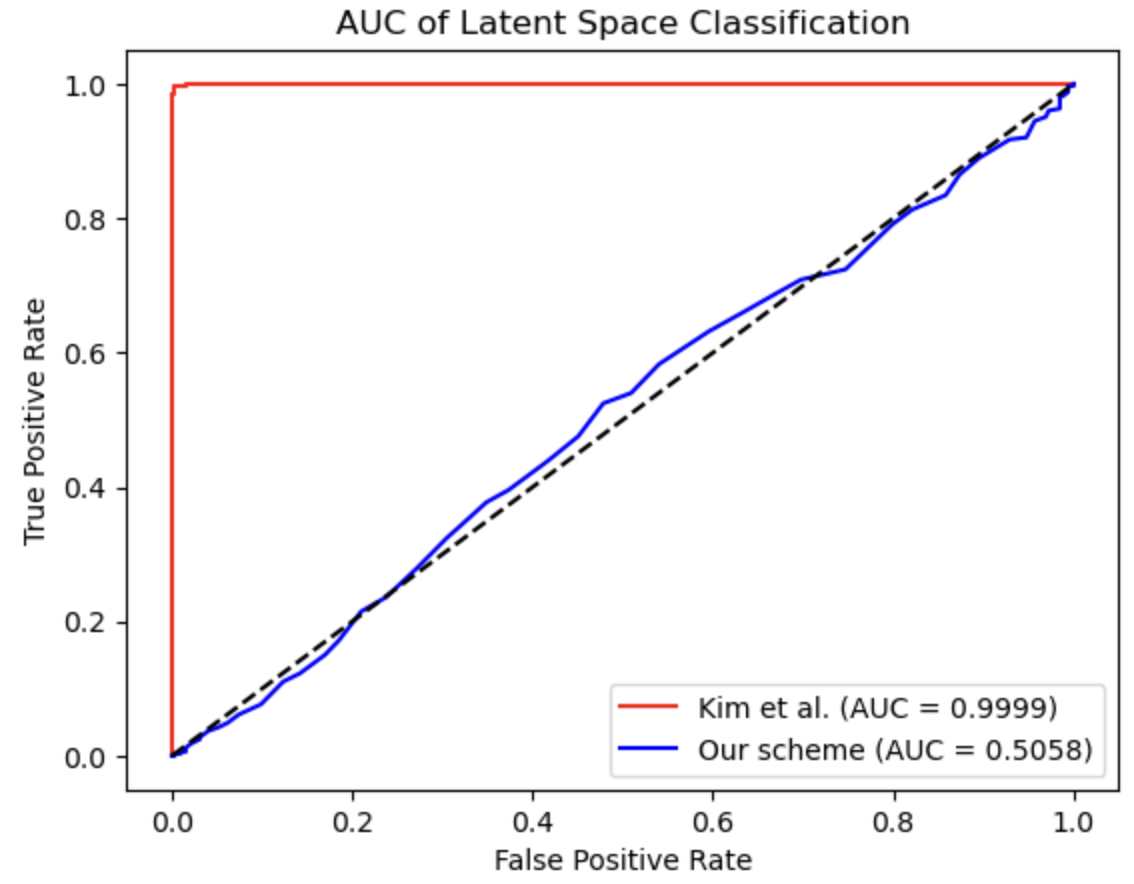}
     \caption{ROC curves and AUCROC for Random Forest classifiers to distinguish latent space with and without embedding. Our method does not alter the latent space and so classifier performance is random guessing. Kim et al.~\cite{kim2023diffusion} (message projection) do alter the distribution, which leads to (almost) perfect classification.}
     \label{fig:latent}

\end{figure}
\section{Security Experiments}

Although we proved the security of our construction, we can nevertheless evaluate security empirically. For this, we can train ML-based distinguishers to identify any distortions caused by the embedding of our messages. We can also evaluate cover (i.e., output) image quality to determine whether embedding covert messages alters the output distribution in a way that an adversary can detect. For this, we use the Fr\'echet Inception Distance (FID)~\cite{heusel2017gans}, CLIP-score~\cite{hessel2021clipscore}, and image classifiers as distinguishers.

\subsection{Security of the Latent Space}
\label{sec:exp:kimetaldist}

We train Random Forest binary classifiers to distinguish latent spaces. Concretely, we generate 3 sets of 500 latent spaces gathered by inverting a cover image. All images are generated and inverted with a DDIM scheduler. The first set does not contain a steganographic message. The second set contains a message embedded using our method, and the final uses a message embedded by (our re-implementation of) the method used by Kim et al.~\cite{kim2023diffusion}. Recall that their embedding method overrides the initial latents by assigning 1 bits to either $\sqrt{2}$ or $-\sqrt{2}$ and 0 bits to 0. 

We create a histogram for each latent space with 10 equally spaced bins ranging from -3 to 3. The bins are filled with the number of feature values that fall within the range of the bins. We then train a RF classifier on the histogram bins distribution counts as features to distinguish between non-steganographic latent spaces, and ones containing a message embedded using our method. We also trained another RF classifier in the same way to distinguish between non-steganographic latent spaces and message projection. We use 400 training samples and 100 test samples for both classifiers.

The results are shown in~\cref{fig:latent}. For our method the classifier only achieves an AUC of 0.5, or random guessing. For Kim et al. it achieves an AUC of 1.0, meaning it was able to identify every single steganographic message without errors. 

\begin{figure}
    \begin{subfigure}[h]{0.495\linewidth}
    \includegraphics[width=\linewidth]{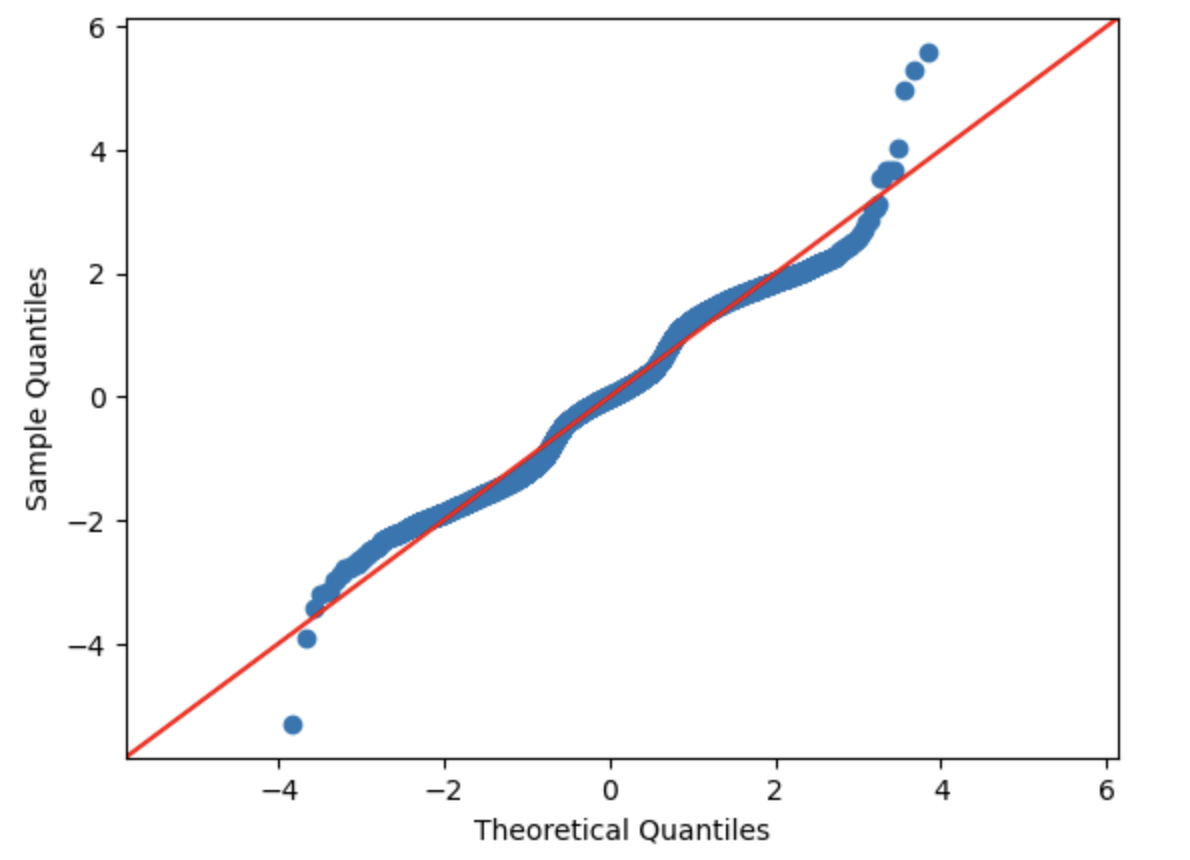}
    \caption{Message Projection}
    \end{subfigure}
\hfill
    \begin{subfigure}[h]{0.495\linewidth}
    \includegraphics[width=\linewidth]{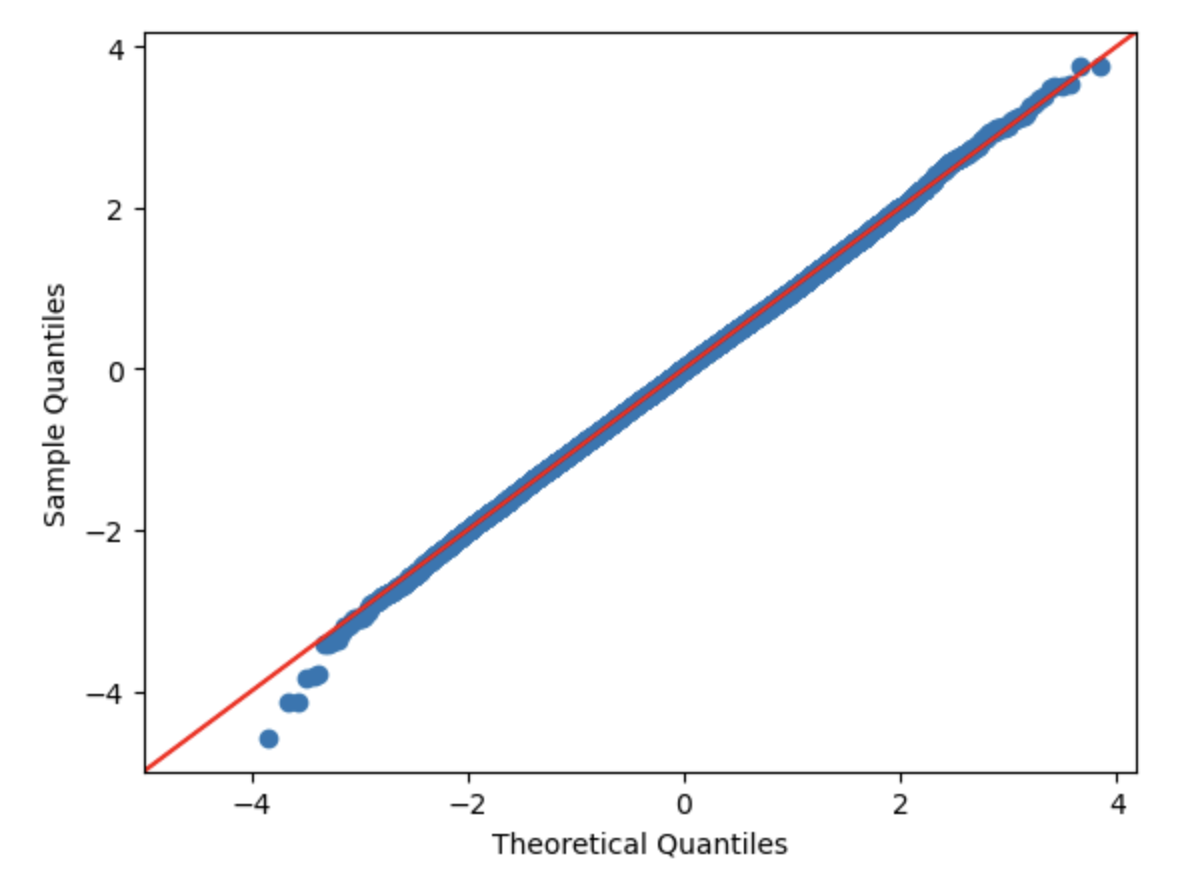}
    \caption{Our Method}
    \end{subfigure}%
\caption{QQ plots of the two embedding methods against the natural recovered latent space distribution. Although the graph on the left follows the natural mean and scale well, the clustering around the values inserted into the feature values is easy to detect.}
\label{fig:qqplot}
\end{figure}

To demonstrate why it is so easy to detect latent space embeddings when they alter the distribution, we also generate QQ-plots from the same sets of latents. First we establish a 'natural' set, which is the distribution across 500 latent spaces recovered from images without an embedded message. We can then compare the two methods. To create the QQ-plot we create a histogram with 100 equally spaced bins ranging from 3 to -3, and then compare the proportion of points within that bin against the natural distribution. We provide the results in~\cref{fig:qqplot}. The results show that simply inserting values according to the bits being embedded leave distinct clusters near the values being used.

To further illustrate the effect of embedding method on returned latent spaces, we performed a two-sample Kolmogorov-Smirnov (KS2) test. We compare the natural set described above against similar sets for our method and for message projection.

Our null hypothesis is that the distributions are the same, and thus the embedding method provides latent space indistinguishability. Using the ks\_2samp function of the scipy.stats package, we compare our method against the natural distribution, which returned a test statistic value of $0.0006$ and $p$-value $0.475$ and thus we do not reject the null hypothesis. For the message projection method, however, the test statistic is $0.055$ with a $p$-value $4.56 \cdot 10^{-49}$ %
and hence the null hypothesis can safely be rejected.

\subsection{Security of the cover}
\label{sec:experiments:coversecurity}

We calculate FID using the Pytorch FID implementation~\mbox{\cite{Seitzer2020FID}}. We generate a large set of images. 10,000 without a message embedded, and 10,000 with a message embedded. To establish a base FID, we split the natural dataset in half, then calculated the FID of the two halves. This returned an FID of 13.688. We then calculated the FID of images with a message embedded within them against images without a message, and got an FID of 13.642.

We also gathered the CLIP-scores~\cite{hessel2021clipscore} of the generated images against the prompt used to generate them. Images without a message had an average CLIP-score of $0.363\pm0.001$. Images with a message had an average score of $0.3628\pm.0008$. Using a t-test with the null hypothesis that these originate from the same distribution, returns a $p$-value of $ 0.1199$, meaning we do not reject the null hypothesis, and these are likely the same distribution.

\begin{table}[ht]
\centering
\caption{Accuracy of various steganalysis models attempting to detect if a generated image contains an embedded message or not. }
\begin{tabular}{cllll}
\toprule
 & CLIP & Xu-net & Ye-net & SiaStegNet \\ \midrule
\begin{tabular}[c]{@{}c@{}}Model Acc\end{tabular} & 0.506 & 0.50 & 0.497 & 0.502\\
\bottomrule
\end{tabular}
\label{tbl:stega}
\end{table}

\begin{table*}[t]

\caption{Metric comparison for similar steganography or watermarking papers. Where multiple methods were present in the paper, we selected the best version. For our results we used metrics for our highest expected value parameters (redundancy $\rho = 6$ and threshold $\tau = 0.3$ and the EDICT scheduler).}
\subfloat[Detectability ranges from 0 to 1 with 0.5 being random guessing. Note: For LSB (Least Significant Bit) steganography the capacity is 1 bit per pixel but the embedding is on the final (full size) image, not in the latent space, therefore the actual total capacity is larger.]{
\begin{tabular}{lllll}
\toprule
 & \begin{tabular}[c]{@{}l@{}}Capacity \\ (BPP)\end{tabular} & \begin{tabular}[c]{@{}l@{}}Bit \\ Accuracy\end{tabular} & Detectability & \begin{tabular}[c]{@{}l@{}}Provable \\ Security\end{tabular} \\ \midrule
Wei et al.\cite{wei2023generative} & 1 & 1 & 0.548 & $\times$ \\
Peng et al.\cite{peng2024ldstega} & 0.4 & 0.995 & Unreported & $\times$ \\
Kim et al.\cite{kim2023diffusion} & 1 & 0.954 & Pe=0.423 & $\times$ \\
LSB & 1 & 1 & 1 & $\times$ \\
This Paper & 0.501 & 0.999 & 0.5 & \checkmark \\ \bottomrule
\end{tabular}
}
\quad
\subfloat[FID and Clip-Scores were measured compared against a natural score without any message embedding. For Tree Ring Watermark, we included the radius of their embedding since they do not provide a capacity measure. ]{
\begin{tabular}{lllll}
\toprule
 & \begin{tabular}[c]{@{}l@{}}Capacity \\ (bits)\end{tabular} & \begin{tabular}[c]{@{}l@{}}Bit \\ Accuracy\end{tabular} & \begin{tabular}[c]{@{}l@{}}FID \\ (Base 25.31)\end{tabular} & \begin{tabular}[c]{@{}l@{}}Clip-Score \\ (Base 0.363) \end{tabular} \\ \midrule

\begin{tabular}[c]{@{}l@{}}Tree Ring \\ Watermark\end{tabular} & r=4 & 0.9999 & 25.47 & 0.356 \\

\begin{tabular}[c]{@{}l@{}}Gaussian \\ Shading\end{tabular} & 256 & 0.9999 & 25.20 & 0.3631 \\
This Paper & 2087 & 0.99 & 25.43 & 0.3628 \\ \bottomrule
\end{tabular}
}

\label{tbl:comparison}
\end{table*}
\subsection{Steganalysis}
\label{steganalysis}
We use these same datasets to train several steganalysis models. Clip~\cite{radford2021learning}, Xu-net \cite{xu2016structural}, SiaStegNet \cite{you2020siamese}, and YeNet \cite{ye2017deep}.
For CLIP we use OpenAI's CLIP model with ViT-B/32 transformer architecture. We create two classes ``steganographic'' and ``non-steganographic'', and attempt to classify images according to them. We use the dataset we created in~\cref{sec:experiments:coversecurity} 10,000 images without a
message embedded, and 10,000 with a message embedded. Note, this is different from Clip-Score where we compared the image against the prompt used to generate them. Here the model is used to extract features on for a ML-based distinguisher. 
Our CLIP model uses an Adam optimizer and measures the cross entropy loss. We train for 30 epochs. The model is unable to distinguish between classes. 

\begin{figure}[!]
     \centering
     \includegraphics[width=0.875\linewidth]{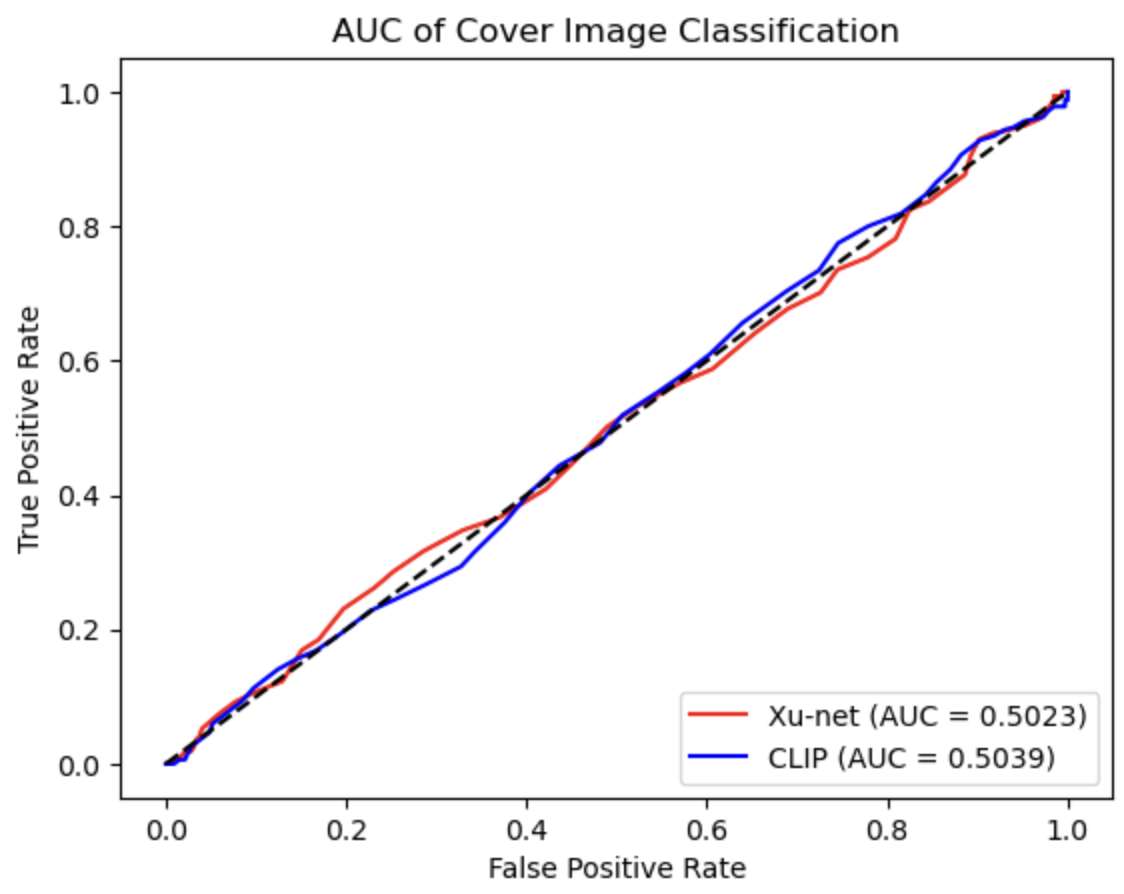}
     
     \caption{Cover Image Classification, showing the CLIP and Xu-net models, both unable to distinguish between generated images with a message embedded using our method vs. without.  Not shown, these models are also unable to detect images generated by Kim et al. }
     \label{fig:output_acc}

\end{figure}

Xu-net, SiaStegNet, and YeNet are originally intended to identify the noise present when embedding an image directly into a cover, and thus require an original cover image to compare against the embedded cover.
This does not exist for our method, so we use cover images generated with the same latent space as our steganographic cover, but with no message embedded in them. Again, the models are unable to distinguish between non-steganographic and steganographic output. The results for CLIP and Xu-net are in \cref{fig:output_acc}. 

We also tested images generated using Kim et al.'s method, and found that none were able to detect messages embedded using their methods. This is consistent with their results, since their method of inserting values maintains enough of the original distribution's shape to maintain image quality. 

\begin{table*}[t]

\caption{Capacity and reliability for EDICT and DDIM schedulers at different redundancy levels, for a threshold of $\tau=0.3$. As well as values at different threshold values for a redundancy of $\rho = 6$. Capacity is the same across both schedulers and calculated in bits per pixel (BBP) at the 64x64 size for consistency with prior work and the latent space. Reliability is the percentage of complete messages correctly recovered.}

\centering
\begin{tabular}{lcccccccccc}
\toprule

 & \multicolumn{5}{c}{Threshold ($\tau=0.3$)} & \multicolumn{5}{c}{Redundancy ($\rho=6$)} \\ \midrule
 & 1 & 2 & 4 & 8 & 16 & 0 & 0.1 & 0.5 & 1 & 1.5 \\
Capacity (BPP) & 3.056 & 1.529 & 0.7639 & 0.3820 & 0.1910 & 0.666 & 0.613 & 0.411 & 0.211 & .089 \\
Reliability (EDICT) & 0\% & 0\% & 7\% & 93.8\% & 100\% & 18.56\% & 36.4\% & 91.0\% & 100\% & 100\% \\
Reliability (DDIM) & 0\% & 0\% & 18.4\% & 86.6\% & 99.6\% & 6.4\% & 20.78\% & 82.4\% & 94.5\% & 100\% \\ \bottomrule
\end{tabular}
\label{table:redun}

\end{table*}



\section{Performance}
\label{sec:perf}
%

We evaluate and discuss tradeoffs between capacity, reliability and processing time. We emphasize that if we had chosen to embed bits in a way that altered the latent distribution, the tradeoffs would also involve detectability. 

\subsection{Metrics}
%
Capacity measures the number of bits that can be communicated in a single cover. Reliability is the measure of how reliable message transmission is; can the sender be sure that their message reaches the receiver? 

We measure capacity in bits per pixel, assuming an image size of 64x64 in order to have standardized measures across models and related work. Different models may transform the latent space into different sized images, but since we are embedding the message into the latent space the final image size is largely irrelevant. 

Reliability can be measured by the likelihood of a message to be received without errors, i.e., it can pass the tag check and be decrypted correctly. To help tune parameters to optimize both of these metrics, we propose a single metric that accounts for both capacity and reliability, called {\em expected bits received}. This is simply the proportion of sent messages that are received in their entirety, times the number of bits that each message contains. Said differently, it measures the amount of bits correctly recovered per message assuming an infinite stream of transmission from the sender to the receiver. By optimizing for this metric, we can ensure that the maximum number of bits make it to the receiver. We propose this metric because optimizing for capacity or reliability alone is a poor strategy. Achieving high capacity is pointless if the messages sent cannot be recovered correctly. 
In Table \mbox{\ref{tbl:comparison}} we show how the metrics of our method compares against similar steganographic and watermarking methods.



\paragraphb{A note about bit accuracy}
Related work typically evaluates reliability as bit accuracy, i.e., the proportion of embedded bits that the receiver correctly recovers. We use this metric in experiments for ease of comparison, but we do not believe it is meaningful. The reason is that when the message is encrypted (as necessary for confidentiality) retrieving any proportion less than all of the embedded bits prevents recovery of the plaintext, since decryption requires all bits of the ciphertext to be correctly recovered. 
Therefore, we advocate for using message accuracy, i.e., proportion of complete plaintext messages correctly recovered, which accounts for encryption.

\subsection{Parameter Tradeoffs}
The three parameters that affect performance the most are redundancy, threshold, and scheduler. We evaluate redundancy for values between 1 (no-redundancy) and 16, and threshold from 0 to 1.5. For the schedulers we use the standard DDIM scheduler and EDICT. In Table \ref{table:redun} we show how the redundancy and threshold settings influence reliability and capacity numbers for both schedulers. The capacity does not depend on the scheduler used, only on the latent space size, redundancy, and threshold. However, reliability varies based on the scheduler as different error correction mechanisms are available in each case. Hence we quantify reliability for both. Unless stated otherwise, we use a base threshold of $\tau = 0.3$, and a base redundancy of $\rho = 6$.

Observe that increasing redundancy improves accuracy since our likelihood ratio can better account for inversion inaccuracies, however it decreases the length of messages that can be sent. The threshold has a similar effect, as when we increase it the likelihood of a component value flipping signs due to inversion decreases improving reliability, however fewer component values pass the threshold decreasing the number of slots available to embed bits. 
\begin{figure}[!th]
    \begin{subfigure}[h]{0.495\linewidth}
    \includegraphics[width=\linewidth]{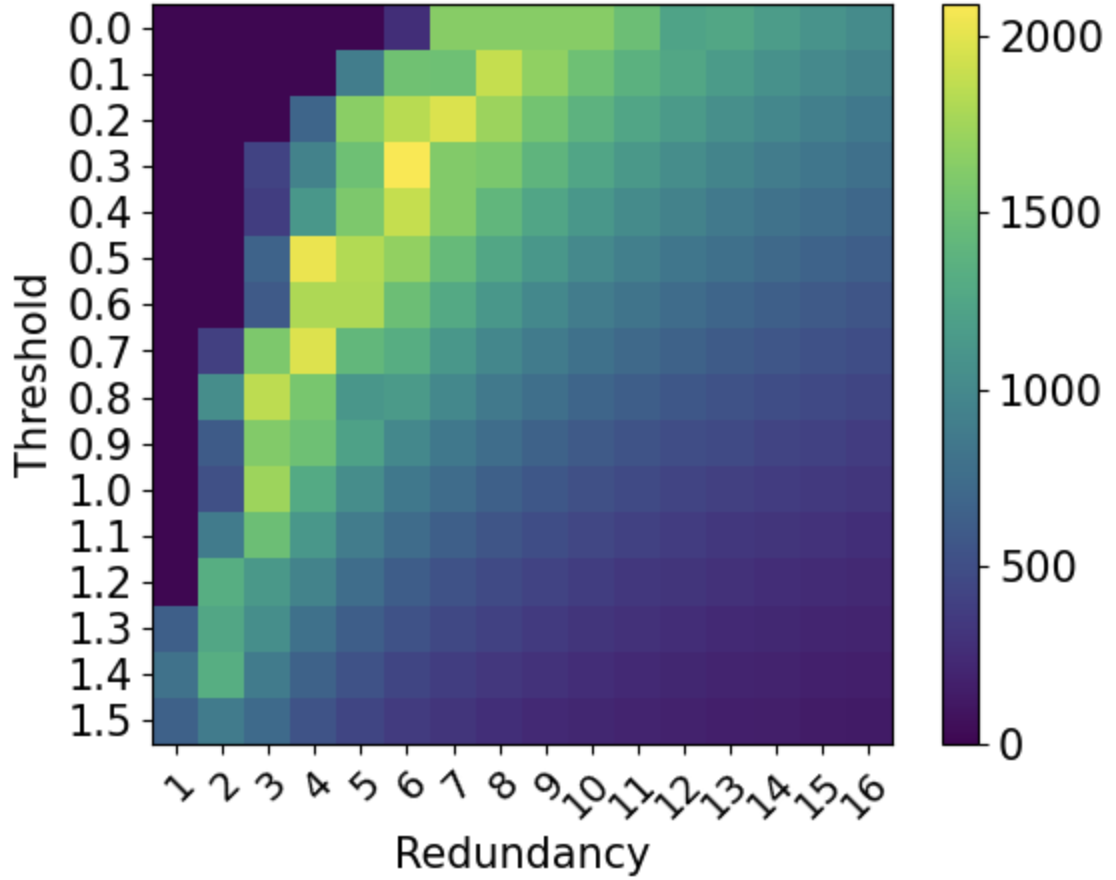}
    \caption{EDICT.\label{fig:heat_edict}}
    \end{subfigure}
\hfill
    \begin{subfigure}[h]{0.495\linewidth}
    \includegraphics[width=\linewidth]{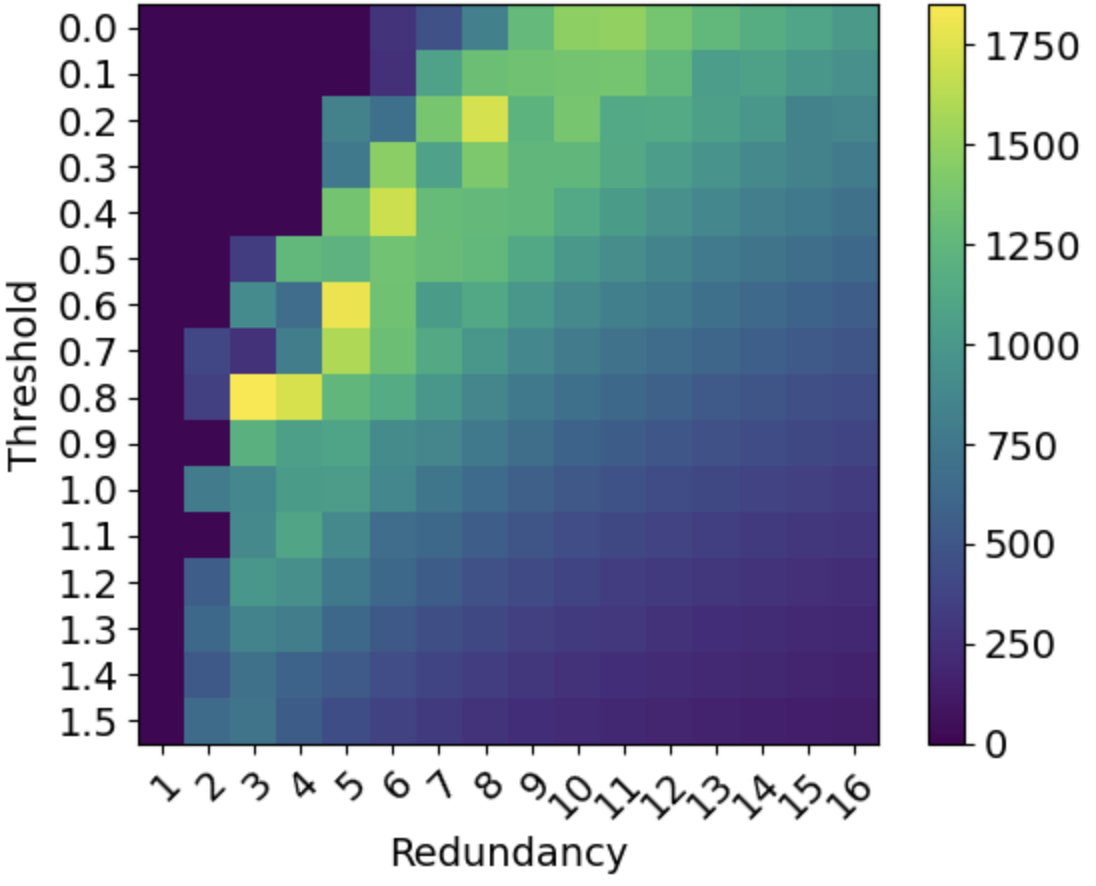}

    \caption{DDIM.\label{fig:heat_ddim}}
    \end{subfigure}%
    
\caption{Heatmap of expected bits received at various thresholds (y-axis) and redundancies (x-axis). }
\label{fig:heat_maps}
\end{figure}

What parameters give the highest expected bits received?

We conducted a grid search over threshold and redundancy levels, shown in~\cref{fig:heat_ddim,fig:heat_edict}, to identify the optimal parameter values. We tested thresholds from $0$ to $1.5$ incrementing by 0.1, and redundancy from 1 to 16 incrementing by 1. We generated 10 messages at each combination of threshold and redundancy. We then calculated expected bits received by multiplying the number of message bits contained in a single image at those parameter values by the probability of the message successfully being received. 

Interestingly there are diminishing returns for both parameters. The optimal setting is $\rho = 6$ and $\tau = 0.3$, which yields an expected bits received of $2087$ (per message).

\subsection{Processing Times}
\label{time}
For some deployment scenario, processing time and latency may be important, so we measure it. For our system, excluding image generation and inversion, sender-side operations take approximately 0.3 seconds, and receiver-side operations take 0.5 seconds. Using StableDiffusion as our underlying model with standard DDIM scheduler, takes approximately 20 seconds to generate an image, and another 20 to invert it. Using EDICT doubles these times, so 40 seconds each for image generation and inversion. 

Further, recall that EDICT error correction uses the dual latents. So depending on the number of discrepancies, or potential errors, between the latents, error correction could take over 10 minutes. For this reason we only attempt to correct errors if there are less than 10 potential errors, which takes about 0.4 seconds on average.

\begin{figure}[th]
    \begin{subfigure}[h]{0.495\linewidth}
    \includegraphics[width=\linewidth]{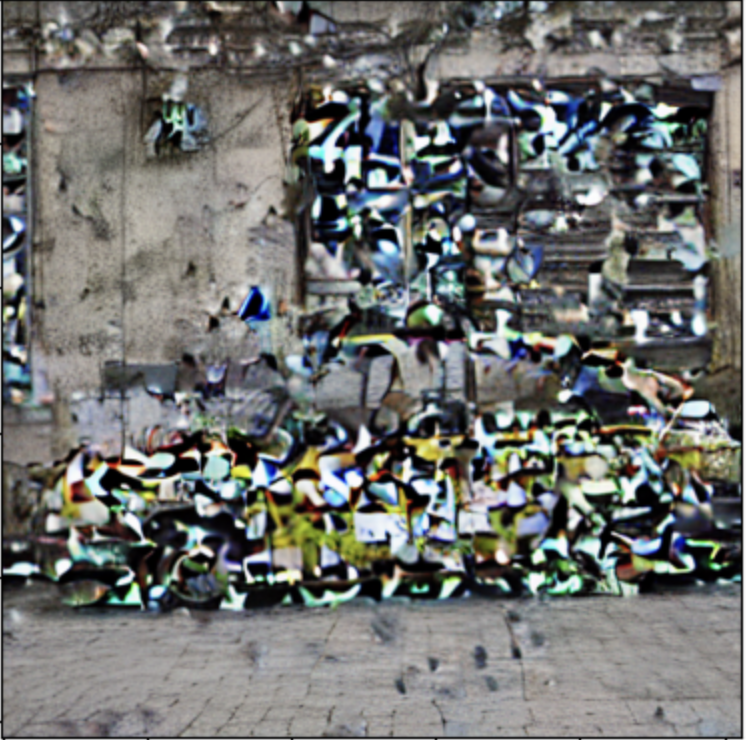}
    \caption{Last Step Embedding}
    \end{subfigure}
\hfill
    \begin{subfigure}[h]{0.495\linewidth}
    \includegraphics[width=\linewidth]{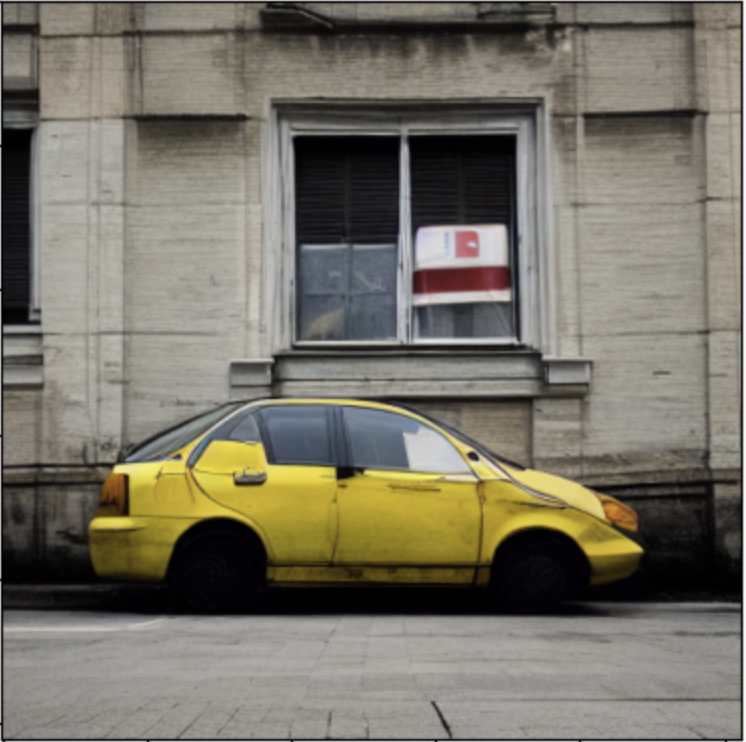}
    \caption{Initial Latent Space Embedding}
    \end{subfigure}%
\caption{These are two images that have the same $\xt$ and embed the same message. Left (a) was embedded into the latent space during the last step of the diffusion process. Right (b) follows our original method of embedding into the original latent space.}
\label{fig:img_comp}
\end{figure}

\begin{table*}[!t]
\centering
\caption{Message accuracy at different levels of transformation, SSIM to show image degradation, and average latent difference. }
\begin{tabular}{ccclcclcclccc}
\toprule
 & \multicolumn{3}{c}{Resize (Downscale)} & \multicolumn{3}{c}{Resize (Upscaling)} & \multicolumn{3}{c}{JPEG Compression} & \multicolumn{3}{c}{Crop} \\
 & 462x462 & 256x256 & \multicolumn{1}{c}{51x51} & 563x563 & 768x768 & \multicolumn{1}{c}{972x972} & 100 & 75 & \multicolumn{1}{c}{50} & 511x511 & 510x510 & 509x509 \\ \midrule
Msg Acc & 0.80 & 0.50 & \multicolumn{1}{l|}{0} & 0.932 & 0.834 & \multicolumn{1}{l|}{0.733} & 0.76 & 0.40 & \multicolumn{1}{l|}{0.166} & 0.7 & 0.365 & 0.034 \\
Bit Acc & 0.999 & 0.992 & \multicolumn{1}{l|}{0.706} & 0.999 & 0.998 & \multicolumn{1}{l|}{0.997} & 0.999 & 0.989 & \multicolumn{1}{l|}{0.966} & 0.998 & 0.994 & 0.981 \\
SSIM & 0.97 & 0.86 & \multicolumn{1}{l|}{0.53} & 0.991 & 0.989 & \multicolumn{1}{l|}{0.980} & 0.999 & 0.962 & \multicolumn{1}{l|}{0.923} & 0.864 & 0.92 & 1.224 \\
Avg Latent Diff & 0.34 & 0.45 & \multicolumn{1}{l|}{0.845} & 0.302 & 0.330 & \multicolumn{1}{l|}{0.366} & 0.311 & 0.466 & \multicolumn{1}{l|}{0.566} & 0.368 & 0.444 & 0.540 \\ \bottomrule
\end{tabular}

\label{tbl:sev}
\end{table*}

\section{Robustness}

It is common for image hosting platforms to resize, crop, compress, or re-encode images. Therefore, it is desirable to consider robustness, which is the ability to recover the message despite distortions to the image. A common problem with steganographic methods that embed into the cover image directly, is that they are very susceptible to distortions~\cite{petitcolas1999information}. When one is attempting to store information inside the least significant bits of a pixel, even minor distortions to the pixel values can undermine reliability. Our premise in this work is that by embedding bits in the initial latents, the deleterious effects of distortions will be much less pronounced. 

To measure robustness we measured reliability for different image formats, and basic image transformations, commonly applied by social media or image hosting sites. We also measured the Average Latent Difference, defined as the average absolute difference between the latents used to generate the image, and that recovered by the inversion process after the transformation/distortion is applied. 

This quantifies how much the transformation affects the latent space, which directly influences our ability to recover the embedded message. To show the effect of the transformation, we also measure SSIM~\cite{wang2004image} between the image before and after the transformation.

To further illustrate the advantage of embedding information in the initial latents, we compare our method against two alternatives. The first alternative is to embed the message in the last latent step $\xts{1}$ before the latent representation is sent through the encoder to become the full size image. This, theoretically, has the advantage of reducing the inversion error since there is little to no noise added between the embedding of the message and the retrieval. However, this process severely impacts the image quality as can be seen in ~\cref{fig:img_comp}. The second alternative is to use a method that embeds bits in the image itself, such as Least Significant Bit (LSB) steganography~\cite{petitcolas1999information}.

In~\cref{tbl:format} we generate 100 messages for each format, converting the image generated by the model into the format listed, then attempt to retrieve the message from the converted cover. We use the default JPG quality (75), but when increased to 100 or 95, it performs similarly to PNG. We also include results for a blur transformation, which blurs the image by setting each pixel to the average value of the pixels in a square box extending radius pixels in each direction. We use radius 1. We add this as an example of an adversarial attack that could be performed without completely destroying the image.

\begin{table}[tbh]
\centering
\caption{Message accuracy, Bit accuracy, average latent difference, SSIM, and accuracy for different embedding methods when switching image formats, and for different transformations.} 

\resizebox{\columnwidth}{!}{%
\begin{tabular}{ccccccc}
\toprule
&
  Msg Acc &
  Bit Acc &
  \multicolumn{1}{c}{\begin{tabular}[c]{@{}c@{}}Avg Latent \\ Diff\end{tabular}} &
  SSIM &
  \multicolumn{1}{c}{$\xts{1}$-Msg Acc} &
  \multicolumn{1}{c}{\begin{tabular}[c]{@{}c@{}}LSB \\ Acc\end{tabular}} \\ \midrule
PNG    & 0.915 & 0.992 & 0.386 & 1     & 0 & 100 \\
JPG    & 0.36  & 0.807 & 1.107 & 0.272 & 0 & 0   \\
TIFF   & 0.865 & 0.993 & 0.313 & 0.96  & 0 & 100 \\
BMP    & 0.885 & 0.991 & 0.449 & 0.94  & 0 & 100 \\ 
Blur   & 0.533 & 0.996 & 0.432 & 0.86  & 0 & 0   \\ \bottomrule

\end{tabular}
\label{tbl:format}
}
\label{tbl:format}

\end{table}

\begin{figure}[t]
     \centering
     \includegraphics[width=\linewidth]{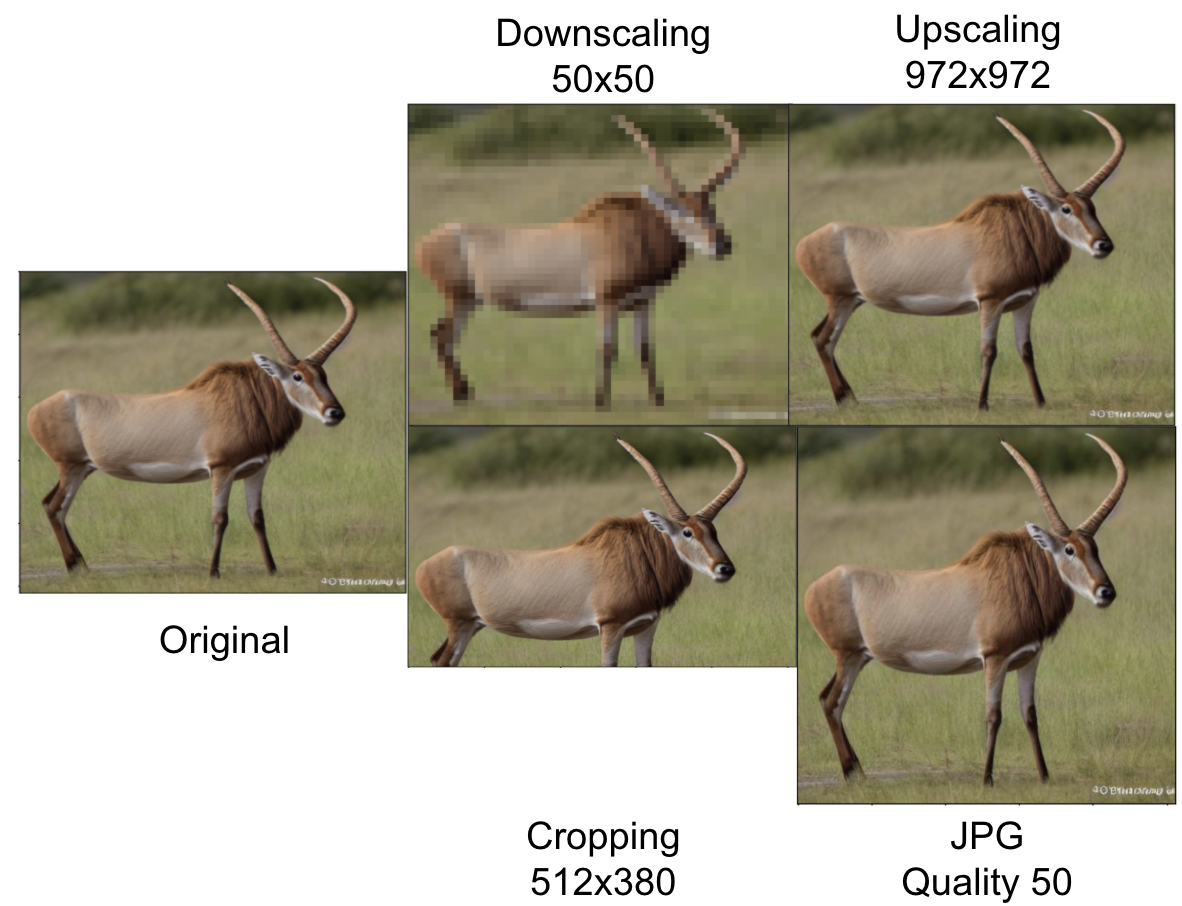}
     \caption{Sample image transformations at high severity.}
     \label{fig:deer}
 \end{figure}

\cref{tbl:sev} shows results for similar tests: 100 samples at the given image transformation level. All distortions are applied using the Python Image Library. JPG is applied by reducing quality to the shown value. For these transformations we use SSIM to establish a range of transformations. We set the minimum to no transformation and the maximum to the point where SSIM stops decreasing, generally when it reaches approximately 0.3. We use this to set 3 levels of transformation. \cref{fig:deer}~shows the image transformations. 

\section{Conclusions \& Future Directions}
We proposed a novel construction to embed covert messages into the output of image-based diffusion models. In the process, we discussed how embedding in the latent space, without disrupting the natural distribution, provides a secure and useful way to perform steganography. 

There remains important avenues for future work. For example, designing an inversion process that further reduces inversion error, would greatly improve usability and robustness. Incorporating ideas from recent research on watermarking could also improve robustness. The security model should be expanded to include user behavior and similarity to other posts within the context. Finally, research into potential misuses of steganographic techniques could be fruitful. 
\section*{Acknowledgments}
We thank the anonymous SaTML reviewers for their helpful comments on earlier versions of the paper. This work was supported in part by the National Science Foundation under CNS-2055123. Any opinions, findings, conclusions, or recommendations expressed in this material are those of the authors and do not necessarily reflect the views of the National Science Foundation.

\bibliographystyle{IEEEtran}
\bibliography{references}

\appendices

\end{document}